\documentclass[english,times]{ettauth}
\pdfoutput=1
\usepackage{babel}
\usepackage[utf8]{inputenc}
\usepackage[OT1]{fontenc}
\usepackage{amsmath}
\usepackage{graphicx}
\usepackage{subfigure}
\usepackage{url}

\def\volumeyear{2009}
\def\DOI{ett.1447}
\def\startpage{504}
\def\endpage{518}
\def\volumeyear{2010}
\def\volumenumber{21}
\gdef\copyrightline{Copyright \copyright\ \volumeyear\ John Wiley \& Sons, Ltd.}

\begin{document}

  \runningheads{M. Rodríguez-Pérez ET AL.}{Persistent congestion problem of FAST-TCP}
  \title{The persistent congestion problem of FAST-TCP: Analysis and solutions}
  \author{M.~Rodríguez-Pérez\corrauth,S.~Herrería-Alonso,
    M.~Fernández-Veiga,C.~López-García}
  \address{Dept.~Telematics Engineering, ETSE~Telecomunicación,
    Campus~Universitario~Lagoas-Marcosende~s/n, 36310~Vigo,~Spain}

  \corraddr{Email address: Miguel.Rodriguez@det.uvigo.es.}
    
  \begin{abstract}
    FAST-TCP achieves better performance than traditional TCP-Reno schemes,
    but unfortunately it is inherently unfair to older connections due to
    wrong estimations of the round-trip propagation delay. 

    This paper presents a model for this anomalous behavior of FAST flows,
    known as the persistent congestion problem. We first develop an elementary
    analysis for a scenario with just two flows, and then build up the general
    case with an arbitrary number of flows. The model correctly quantifies how
    much unfairness shows up among the different connections, confirming
    experimental observations made by several previous studies.

    We built on this model to develop an algorithm to obtain a good estimate
    of the propagation delay for FAST-TCP that enables to achieve fairness
    between aged and new connections while preserving the high throughput and
    low buffer occupancy of the original protocol. Furthermore, our proposal
    only requires a modification of the sender host, avoiding the need to
    upgrade the intermediate routers in any way.
  \end{abstract}


\maketitle


\section{INTRODUCTION}
\label{sec:introduction}

Delay-based congestion avoidance (DCA) algorithms are a promising alternative
to standard congestion avoidance algorithms which employ packet loss as an
indicator of network congestion~\cite{wei06:_fast_tcp,jain89,brakmo94tcp}. In
fact, DCA algorithms outperform TCP-Reno~\cite{jabocson90:reno} and its
variants~\cite{floyd99:_new_reno_modif_fast_recov_algor} in the aspects of
overall network utilization, stability and low buffer occupancy
\cite{wei06:_fast_tcp,brakmo94tcp,martin03}. 

FAST-TCP~\cite{wei06:_fast_tcp} is a good example of pure DCA methods. This
algorithm reacts to increments in round-trip time (RTT) in an attempt to avoid
network congestion before losses occur. Though it achieves higher throughput,
lower transit delays and fewer packet losses than previous versions of TCP,
the FAST congestion avoidance algorithm exhibits some anomalous behaviors that
lead to an inefficient or unfair use of network resources.
It may be regarded as its less harmful effect that the performance degrades
when congestion arises in the return path, because several enhancements have
been recently proposed that can overcome such
problems~\cite{fu03:_remed_for_perfor_degrad_of,chan03:_enhan_conges_avoid_mechan_for_tcp_vegas,liu05:_enhan_tcp_vegas_for_asymm_networ,herreria07}.

A second weakness is put in appearance when flows from different TCP
implementations share a link and compete for bandwidth. Mixed with Reno or
alike versions, in which packet drops are the unique congestion signals the
sender reacts to, FAST flows are unable to achieve their fair share of
bandwidth~\cite{bonal99:_compar_of_tcp_reno_and_tcp_vegasy,weigle06:_perfor_compl_high_speed_tcp_flows}
for the fundamental reason that their implicit utility functions are
different~\cite{Low02,Kunniyur03}, thus making the network have multiple
operating points~\cite{tang05:_equil_and_fairn_of_networ}. Nevertheless, this
phenomenon does not affect the stability of the network, neither does it
affect intra-protocol fairness, nor does it prevent FAST from being useful in
homogeneous high-throughput network domains. Furthermore, there are proposals
that try to detect~\cite{rodriguez08:_heuris_approac_to_passiv_detec} or even
react to this condition achieving a fair share of the bandwidth in the long
run~\cite{king05:_tcp_afric,tang06:_heter_conges_contr}.

The third idiosyncratic behavior of FAST is known as the \emph{persistent
  congestion problem}~\cite{wei06:_fast_tcp,hengartner00:_tcp_vegas_revis},
and is a side effect of the procedure for detecting congestion. Recall that
FAST interprets the increments in the RTT as a sign of incipient congestion.
For tracking those changes, it keeps both an accurate measure of the current
RTT and an estimate of the round trip propagation delay, which is equated to
the minimum RTT observed throughout a single connection. This sampling renders
an almost exact value when there is actually a single FAST flow in the
network, or when the bottleneck link is shared with TCP-Reno flows. In the
second case because the buffer at the bottleneck is likely to empty
frequently. However, when a set of FAST flows coexist in the network, the
dynamics of the congestion control algorithm lead the buffer occupancy along
their paths approximately to a constant level. At this point, if a new
connection starts, it would mistake the current RTT for the propagation delay.
This overestimation is the cause of serious unfairness among the contending
flows~\cite{hengartner00:_tcp_vegas_revis,trinh08:_revis_fast_tcp_fairn,mo99:_analy_compar_tcp_reno_vegas}.


Several techniques have been proposed to correct persistent congestion and
ameliorate fairness between new and aged FAST
connections~\cite{Low02,la99,athur01,chan04,tan05}. Unfortunately, all these
solutions rely on queue management mechanisms at the intermediate routers,
thus hindering large scale deployment. A sound tentative solution to solve
persistent congestion without assistance from routers has been presented
in~\cite{cui06}. It relies on a new method to obtain a good estimate of the
propagation delay that consists, basically, in throttling down briefly each
newly started flow to allow router queues to empty. Nevertheless, as we
demonstrate in this paper, this approach is not effective in all
circumstances.

We provide a mathematical model that predicts the buffer occupancy and the
individual throughput for unsynchronized FAST connections. The results of this
analysis are used to clarify the network conditions that lead to the flawed
performance of the rate reduction method. Moreover, this model is also used to
build a solution able to remove the undesired effect of persistent congestion
under more general conditions. As in~\cite{cui06}, our proposal only requires
the modification of the sender end host and, consequently, router queues can
remain simple FIFO buffers without jeopardizing FAST performance. Further, our
solution retains the useful properties of FAST-TCP attaining high throughput
and low router buffer
utilization~\cite{rodriguez08:_achiev_fair_networ_equil_with}.

The rest of this paper is organized as follows. Section~\ref{sec:related-work}
summarizes previous work on models for the FAST-TCP behavior and persistent
congestion issues. Section~\ref{sec:background} gives a brief overview of pure
DCA algorithms in general and FAST-TCP in particular. In
Section~\ref{sec:analytical-model}, we analyze the persistent congestion
problem and its impact on the transmission rates of FAST flows on several
scenarios. Section~\ref{sec:rate-reduction} illustrates why the rate reduction
method proposed in~\cite{cui06} fails to solve this bias in networks with
small propagation delays or when they are shared by many flows. In
Section~\ref{sec:our-prop}, we present a solution to the persistent congestion
problem that lacks the limitations of the rate reduction method.
Section~\ref{sec:exper-valid} contains some simulation experiments that
validate both the proposed analysis and our solution. Lastly,
Section~\ref{sec:conclusions} summarizes this work.


\section{RELATED WORK}
\label{sec:related-work}

Because of their ability to keep the network out of congestion, delay based
congestion avoidance (DCA) proposals started to appear in the literature
shortly after the appearance of Van Jacobson's seminal paper of TCP congestion
control~\cite{jacobson88congestion}. In fact, in~\cite{jain89} a proposal is
presented to use delay for congestion avoidance in interconnected networks.
However, the first approach to complement TCP congestion control with a delay
avoidance algorithm did not appear until the TCP-Vegas
proposal~\cite{brakmo94tcp}. Models for the behavior of Vegas started
appearing shortly after. Most of them are also applicable to other pure-DCA
approaches, that is, protocols that only need packet delay measures to react
to congestion, as, for instance, FAST-TCP~\cite{wei06:_fast_tcp}.

These first works focused on the interactions between pure DCA approaches and
the predominant TCP-Reno congestion control
algorithm~\cite{bonal99:_compar_of_tcp_reno_and_tcp_vegasy,mo99:_analy_compar_tcp_reno_vegas,hengartner00:_tcp_vegas_revis},
and found both approaches not entirely compatible. In particular, Reno based
algorithms were shown to be more aggressive occupying network buffers than DCA
flows. This, coupled with a FIFO queueing discipline, made the latter get a
throughput lower than expected. However, in the recent years some proposals
have appeared that try to mitigate this problem by providing methods for pure
DCA algorithms to self-tune themselves for being more aggressive when
competing against Reno-like
traffic~\cite{king05:_tcp_afric,tang06:_heter_conges_contr}. The problem with
these proposals is that they either are very slow reacting or that they behave
like Reno when the network is congested.
In~\cite{rodriguez08:_heuris_approac_to_passiv_detec} a proposal is made to
better detect the presence of Reno-like traffic that could lead to more rapid
reaction to its presence, finally giving an incentive to users to employ FAST
congestion control without having to sacrifice performance.

Another intrinsic limitation of some pure DCA methods, and the one that this
paper deals with, is the \emph{persistent congestion} problem. It was first
described for Vegas flows in~\cite{hengartner00:_tcp_vegas_revis} and it is a
direct consequence of an important property of both Vegas and FAST, namely,
that once equilibrium is reached, they maintain a constant amount of traffic
queued at network routers. Although this is usually a nice property, as it
prevents jitter in packet arrivals and keeps latency to a minimum, it also
hinders the measurement of the propagation delay. DCA protocols estimate the
round trip propagation delay as the smallest measure of the round trip time
during a whole connection. When the bottlenecks empty frequently, like it is
the case when the network is shared with Reno flows, this procedure can
produce accurate measurements. This is precisely the reason why the LEDBAT
algorithm~\cite{shalunov09:_low_extra_delay_backg_trans_ledbat} does not
encounter the persistent congestion problem in current networks. However, when
the network is being used by pure DCA flows, bottlenecks always contain a
certain amount of enqueued traffic and thus the propagation delay is
overestimated. This overestimation, or more exactly, the different estimations
done by the different flows depending on the minimum network load during their
lifetime, is the cause of serious unfairness. Non pure DCA algorithms, like,
for instance, Compound TCP~\cite{tan06:_compoun_tcp_approac_for_high} do not
suffer from this problem, because the non DCA component of the protocol
guaranties that the buffer occupation varies with time. Thus, the estimation
of the round trip propagation delay improves as successive measures of the
round trip time are more likely to encounter emptier buffer at bottlenecks
after packet losses.

To illustrate it, we have run a short experiment with the help of the
ns-2~\cite{ns-2} simulator. In a simple dumb-bell topology we have set up two
kind of FAST flows sharing a gigabit bottleneck. Flows of the first kind,
\emph{long flows}, send $500\,$megabytes of data, while the second kind of
flows, \emph{short flows}, only send $20\,$megabytes. Flows arrive at the
network following two independent Poisson arrival processes.
\begin{figure}
  \centering
  \includegraphics[width=\columnwidth]{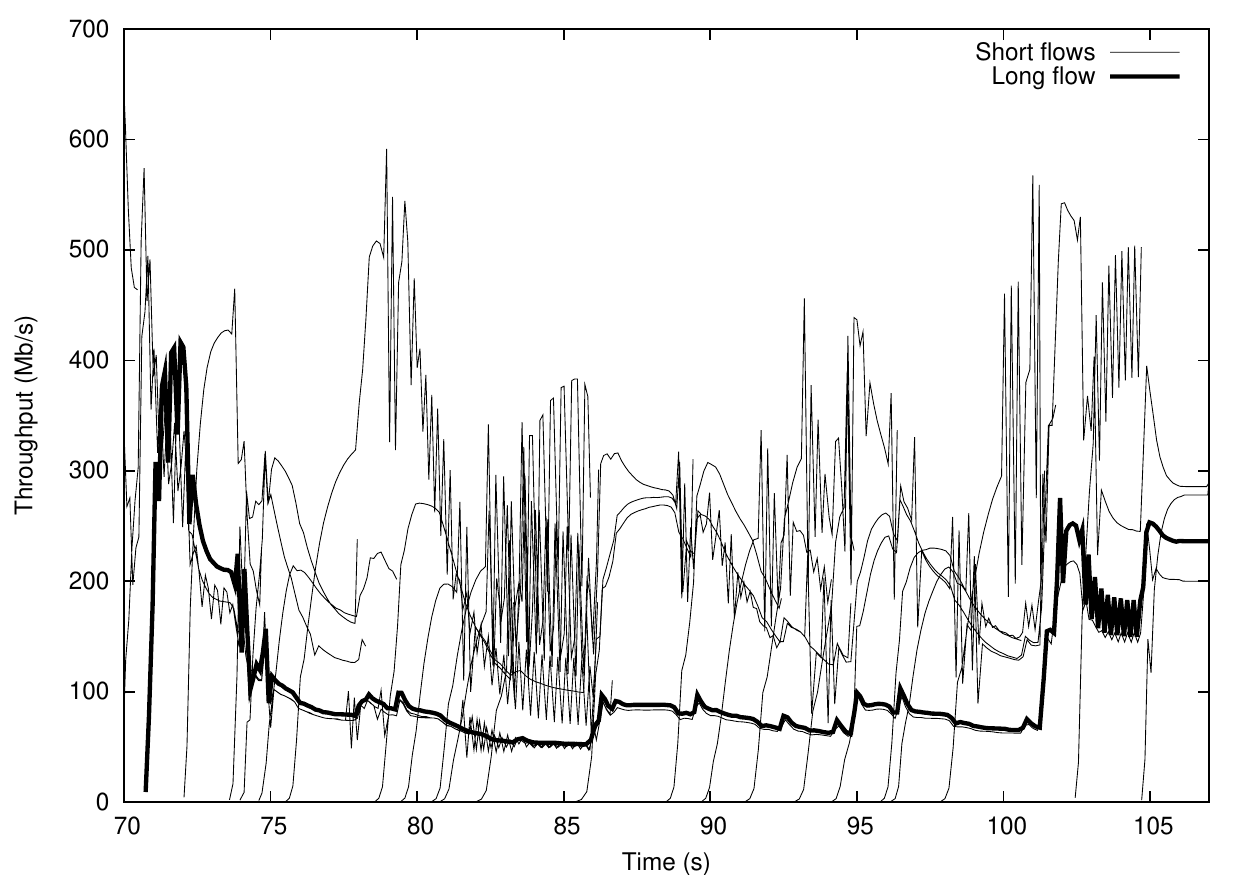}
  \caption{Example trajectory of long running FAST flows \textrm{vs} short
    flows showing the persistent congestion effect.}
  \label{fig:persistent-example}
\end{figure}
In Fig.~\ref{fig:persistent-example} we can see in detail a small part of the
simulation. The thicker line represents the throughput of a paradigmatic long
flow. We see how at the start it takes more share than its fair rate, however,
as time passes and shorter flows arrive at the network they consistently
obtain more share than the long flow. Thus, as long flows suffer the arrival
of more flows in their lifetime than short flows, they suffer more the
consequences of the persistent congestion problem and are unable to get their
fair share of network resources. Moreover, short flows enjoy for a significant
percentage of their lifetime being the last coming flow to the bottleneck, and
thus they benefit from the persistent congestion problem. In fact, we have
measured in the above scenario that long flows take on average $40\,\%$ longer
to finish than what it would take if they had enjoyed their fair share.

The first mathematical model of this problem appeared in~\cite{Low02}, which
provided a recursive set of equations to calculate the throughput of a set of
flows reaching a bottleneck consecutively. However, this model employs a
different interpretation of Vegas parameters than those found in actual
implementations. This further biases the results against old flows.

In this paper we will provide a set of equations according to the actual
interpretation of FAST (and Vegas) parameters for a scenario of sequential
pure DCA arrivals to a single bottleneck and for the arrival of a single pure
DCA flow to a bottleneck that is already being fairly shared. This latter case
will help us to provide a better mechanism to estimate the propagation delay
and thus fix the persistent congestion problem.

We build on equations for the expected throughput of a pure DCA connection in
equilibrium provided also by these previous works. In fact, much of the work
on this paper builds on the equilibrium formulas found
in~\cite{samios03:_model_throug_of_tcp_vegas} for Vegas
and~\cite{wei06:_fast_tcp} for FAST, that, in the absence of heavy congestion
(packet losses) happen to be identical.

Different improvements have been proposed to correct persistent congestion and
ameliorate fairness between new and aged pure DCA connections. \cite{la99}
proposes the use of RED gateways at the routers to get a more even
distribution of the bandwidth regardless of the starting time of competing
flows. However, finding the appropriate threshold values for the RED gateways
is not an easy problem and remains an open issue~\cite{alemu04}. \cite{Low02}
suggests a way to eliminate persistent congestion using REM at the routers.
REM~\cite{athur01} is an active queue management scheme that keeps buffer low
while leading to the high utilization of the link at the same time. Certainly,
with small queues, the minimum of all measured RTTs is an accurate
approximation to propagation delay. In~\cite{chan04} a new IP option named AQT
(Accumulate Queueing Time) is defined to collect the queueing time experienced
by FAST packets along the path. With this scheme, FAST sources must send some
probing packets with the AQT option active while routers must compute the
queueing time for each receiving probing packet and add it to the AQT field.
As a result, each connection is able to obtain a good estimate of the
propagation delay canceling out the queueing time from the RTT measurement.
\cite{tan05} solves the persistent congestion problem by marking the ToS field
in the IP header with the highest priority for the first packet of each flow.
With priority queueing at routers, highest priority packets will be dispatched
immediately even if the router buffer is not empty and, therefore, FAST-TCP
will obtain an accurate estimate of the propagation delay.

Unfortunately, all these solutions rely on queue management mechanisms at the
intermediate routers, thus hindering large scale deployment. An interesting
method to obtain a good estimate of the propagation delay without assistance
from routers has been presented in~\cite{cui06}. Basically, it consists on
throttling briefly each newly started flow in an attempt to empty router
queues so that it can obtain a good estimate of its true propagation delay.
Nevertheless, as we will demonstrate in Section~\ref{sec:rate-reduction}, this
solution is not effective in networks with short propagation delays or when
shared by many flows.


\section{FAST-TCP DESCRIPTION}
\label{sec:background}

DCA algorithms work on the assumption that it is possible to gain insight into
network status by observing the variations in the RTT. The difference between
the RTT and the propagation delay is directly related to the amount of data in
transit, and, the more data in the network, the nearer it is to become
congested. So, adjusting the window size based on these variations, DCA flows
keep an appropriate transmission rate without causing congestion. In contrast,
TCP-Reno and its variants need to drive the network to congestion to receive
the feedback needed to adjust the window. They keep slowly incrementing the
window size until it reaches a point that buffers overflow and packet losses
occur. This leads sources to abruptly reduce their sending rates and the slow
increment phase begins again, preventing Reno flows from fully using all the
available bandwidth. Thus DCA algorithms are more suitable in long fat pipes
where packet losses are too scarce to properly adjust the rate or for those
applications negatively affected by sudden changes in the transmission rate.

Both TCP-Vegas and FAST-TCP employ a similar \emph{modus operandi}, in fact,
FAST-TCP can be treated as an improved (faster) version of the
former~\cite{wei06:_fast_tcp}. Throughout the rest of this paper we will focus
on FAST, although most results can also be applied to Vegas with minor or no
adjustments.

To modulate its transmission rate FAST-TCP employs a congestion window
analogous to the one employed by Reno variants. The FAST congestion window can
be characterized, at the flow level, by the following dynamic equation:
\begin{equation}
  \label{eq:dynamic_fast}
  \dot w(t) = \gamma \alpha \left(
    1 - \frac{q(t) x(t)}{\alpha}
    \right),
\end{equation}
where $\gamma$ and $\alpha$ are configuration parameters, $q(t)$ is the
instantaneous queueing delay and $x(t)=\frac{w(t)}{d+q(t)}$ is the
transmission rate, $d$ being the round trip propagation delay. This equation
has the property that the variation on the window size is directly
proportional to the distance from equilibrium, yielding very \emph{fast}
convergence times.

The above dynamic flow level behavior is implemented at the packet level with
the following rule. Every update interval, defined to be a constant time or
some number of RTTs depending on the precise FAST version, the window size is
updated as
\begin{equation}
  w \longleftarrow \gamma \left( \frac{\hat d w}{\hat r} + \alpha \right) +
  (1-\gamma)w,
  \label{eq:wupdate}
\end{equation}
where $\hat d$ is the current estimation of the round trip propagation delay
and $\hat r=\hat d + \hat q$ is an estimation of the round trip time. The
accurate estimation of $d$ is a bit tricky as it can only be correctly
measured in the absence of cross traffic. In practice $\hat d$ is set to the
minimum round trip time observed during the whole transmission. In the end,
this is only a problem when different FAST flows have different
overestimations. As long as all FAST flows make the same error, the fairness
properties are not affected. Later in this paper, we will study this problem
and provide solutions for it. Eq.~\eqref{eq:wupdate} brings light to the
meaning of $\gamma$. In fact it is just a \emph{smoothing factor} or
\emph{gain} that controls the speed of convergence. Its value its taken from
the semi-open range $(0,1]$, although $\gamma=0.5$ is its more common value.

The equilibrium properties of FAST are well established in the
literature~\cite{wei06:_fast_tcp,jin03:_fast_tcp} and coincide with those of
Vegas~\cite{Low02,samios03:_model_throug_of_tcp_vegas}. In fact, they are a
direct consequence of its dynamic equation~\eqref{eq:dynamic_fast}. It
suffices to make $\dot w(t)=0$ to obtain that under equilibrium each
connection attains a throughput
\begin{equation}
  \label{eq:fast-tput-equilibrium}
  x^* = \frac{\alpha}{q^*} = \frac{\alpha}{\hat r^* - \hat d}.
\end{equation}
This equilibrium formula merits some observations. Firstly, as long as all
FAST flows share the same configuration (same $\alpha$ value) and bottleneck,
they all obtain the same throughput, irrespective of their propagation delays.
Secondly, it helps us to get insight into what is the effect of the $\alpha$
parameter. In fact, if all the flows sharing a bottleneck, and thus observing
the same queueing delay ($\hat r^* - \hat d$), are set up with the same
$\alpha$ parameter, they will get the same equilibrium transmission rate
$x^*_i = x^*$. It then follows that
\begin{equation}
  \label{eq:enqueued-packets}
  (\hat r^* - \hat d)\sum_{i=1}^n x_i^* = (\hat r^* - \hat d)C = n\alpha,
\end{equation}
for $n$ flows and link capacity $C$. Finally, solving for $\alpha$
in~\eqref{eq:enqueued-packets} it becomes apparent that each flow contributes
$\alpha$ packets to the bottleneck backlog.

There is a trade off for selecting an appropriate $\alpha$ value. On the one
hand we want to select a small value, to minimize overall latency and buffering
needs in the network. On the other hand, big values provide faster convergence
times. At the same time, too small $\alpha$ values can produce too little
queueing delays making their precise measure too difficult for end hosts.

Lastly, it must also be taken into consideration that not every combination of
$\alpha$ and $\gamma$ produces stable configurations. Several papers deal with
the conditions that both $\alpha$ and $\gamma$ must meet to reach equilibrium.
More details can be found
in~\cite{wang05:_model_and_stabil_of_fast_tcp,choi05:_global_stabil_of_fast_tcp,choi06:_global_expon_stabil_of_fast_tcp,tan07:_param_tunin_for_fast_tcp}.


\section{ANALYTICAL MODEL}
\label{sec:analytical-model}

Throughout this section we will build on the model for permanent congestion
developed in~\cite{Low02} with the necessary adaptations for FAST and with a
focus on reaching a closed-form formula that predicts persistent congesting
effects. So, we will consider a stable all-FAST scenario where new flows
arrive and modify the equilibrium throughput.

Under these conditions, with each flow keeping its amount of enqueued data in
the network never below $\alpha$ packets, new FAST flows are unable to obtain
accurate measures of the path delays. Please note that scenarios with other
types of TCP flows are of no interest to our study, as router queues get
eventually empty, giving a chance to the FAST flows to accurately estimate
their propagation delay.

\subsection{Two flows scenario}
\label{sec:two-flows}

We present firstly the most elementary case with just two FAST flows appearing
consecutively on a network and sharing a single bottleneck. Without loss of
generality, let us assume that $d_i$ is the actual propagation delay of flow
$i$, and that flow $j>i$ starts after flow $i$ reaches equilibrium.

Recall from Section~\ref{sec:background} that in an all-FAST scenario without
losses caused by congestion, each flow $i$ achieves at equilibrium a throughput
\begin{equation}
  \label{eq:vegas-tput}
  x^*_i = \frac{\alpha_i}{\hat r^*_i - \hat d_i}.
\end{equation}
where $\hat d$ is an estimate of the round-trip propagation delay and $r$ is
the current RTT.\footnote{This equilibrium throughput is not exclusive of
  FAST-TCP, but is also obtained by at least by
  TCP-Vegas~\cite{samios03:_model_throug_of_tcp_vegas}, in an all-Vegas
  scenario.} Let $\alpha_i = \alpha$ for each flow $i$ so as to achieve
fairness~\cite{hasegawa99:_fairn_and_stabil_of_conges}. Then, for the first
flow where $\hat d_1 = d_1 + \sum_l t_{\mathrm{tx}_l}$, $t_{\mathrm{tx}_l}$ is
the transmission time of a packet on the $l^\mathrm{th}$ link, and $r^*_1 =
d_1 + \sum_l t_{\mathrm{tx}_l} + \alpha / C$,
\begin{equation}
  \label{eq:tput-1flow}
  x^*_1 = \frac{\alpha}{(d_1 + \sum_l t_{\mathrm{tx}_l} + \alpha / C) - 
    (d_1 + \sum_l t_{\mathrm{tx}_l})} = C,
\end{equation}
where $C$ is the capacity of the bottleneck link expressed in packets per
second.

After the second flow starts, $\alpha$ more packets should get enqueued at the
bottleneck, and thus the RTT of the first flow would increase by $\alpha / C$,
yielding hypothetical values of $\hat r'_1 = \hat d_1 + 2\cdot\alpha/ C$ and $x'_1 = C
/ 2$, as expected for a fair share of the bottleneck bandwidth. However, the
second flow measures a wrong value for the propagation delay, because it
encounters $\alpha$ packets already enqueued at the bottleneck when it starts.
Since the buffer occupancy never decreases, this causes an overestimation of the
propagation delay and $\hat d'_2 = d_2 + \sum_l t_{\mathrm{tx}_l} +
\alpha/C=\hat r^*_1$. If
this second flow were to enqueue just $\alpha$ packets in the network, its RTT
would then become $\hat r'_2=\alpha/C+\hat d'_2 = 2\cdot \alpha/C+d_2+\sum_l
t_{\mathrm{tx}_l}$, for a (hypothetical) throughput of
\begin{equation}
  \label{eq:tput-2flow}
  x'_2 = \frac{\alpha}{\alpha / C + \hat d'_2 - \hat d'_2} = C.
\end{equation}
However, in the course of reaching $x'_2$, the aggregate throughput surpasses
the bottleneck capacity, making the queue grow and leading to a bigger
queueing delay. Let us call $r^{\tau}_1(t)$ and $r^{\tau}_2(t)$ the new round
trip time measured by the first and the second flow after the second flow has
enqueued at least $\alpha$ packets in the network. Because both flows see the
same increase in queueing delay, we can write $r^{\tau}_1(t)=\hat
r'_1+\frac{\delta(t)}{C} = \hat r^*_1 + \frac{\alpha + \delta(t)}{C}$ and
$r^{\tau}_2(t)=\hat r'_2+\frac{\delta(t)}{C}$, where $\delta(t)$ accounts for
the increase in queue length because of permanent congestion. For notational
simplicity let us define $a(t)$ such that $\delta(t)=\alpha\cdot a(t)$, and lets
call $r^{\tau}_i$, $\delta$ and $a$ the values of, respectively,
$r^{\tau}_i(t)$, $\delta(t)$ and $a(t)$ after equilibrium is reached again.
With this notation, the throughput of each flow in the new equilibrium can be
written as
\begin{eqnarray}
  \label{eq:tput_equil_2fast_first}
  x'_1 &= \frac{\alpha}{r^{\tau}_1-\hat d^*_1} = \frac{\alpha}{r^*_1 +
    \frac{\alpha(1+a)}{C} - \hat d^*_1} = \frac{C}{2+a},\\
  \label{eq:tput_equil_2fast_second}
  x'_2 &= \frac{\alpha}{r^{\tau}_2-\hat d'_2} = \frac{\alpha}{r'_2 +
    \frac{\alpha a}{C} - \hat d'_2} = \frac{C}{1+a},
\end{eqnarray}
taking into consideration that $r^*_1 - \hat d^*_1 = r'_2 - \hat d'_2 =
\alpha/C$.

The new equilibrium will be reached when the two following conditions are met:
\begin{enumerate}
\item Both flows notice that they have enqueued $\alpha$ packets in the
  network, and
\item The aggregated throughput does not exceed the link capacity.
\end{enumerate}
That is, $x'_1+x'_2=C$, and
solving for $a$ yields
\begin{equation}
\label{eq:atwoflows}
a = \frac{\sqrt{5}-1}{2}.
\end{equation}
For this particular $a$ value, it holds that $x'_1 = (1-a)\cdot C$ and $x'_2 =
a\cdot C$. That is, for the case with just two flows, the unfair bottleneck
share achieved after both of them reach equilibrium is independent of $\alpha$
and the respective propagation delays of the flows.

\subsection{Several flows arriving consecutively}
\label{sec:many-flows}

We now extend the previous model to a scenario with an arbitrary number of
flows $n$ arriving sequentially. This is the same case studied
in~\cite{Low02}.

In our new scenario with $n$ flows, flow $i$ starts after the flow $i-1$
stabilizes. Extending the reasoning of the previous version, we can define
$\hat r^{\tau}_i(t) = \hat d^*_i + \frac{(n-i)\alpha + \delta(t)}{C}$, as flow
$i$ only \emph{sees} in its queueing delay the traffic enqueued by flows
arriving later.

Under this condition, and without using any proposal to avoid
persistent congestion, the normalized throughput of the $i^\mathrm{th}$ flow
is
\begin{equation}
  \label{eq:n-flows-rtt}
  \frac{x'_i}{C} = \frac{1}{1+ n-i + \sum_{j=i}^n a_j},
\end{equation}
where $a_j$ accounts for the increase in the queue size required to make
$\sum_{i=1}^j x'_i /C \leq 1$ after the $j^\mathrm{th}$ flow joins. Obviously,
$a_0 = 0$. Notice how, again, the values of $a_i$ and, hence, the relative
throughput obtained by each flow, do not depend on $C$ or any other network
property.\footnote{Note that eq.~\eqref{eq:n-flows-rtt} predicts better
  fairness than~\cite{Low02}. This is because the latter uses an
  interpretation of Vegas behavior that further biases results towards flows
  with larger propagation delays. They consider $\alpha$ and $\beta$
  parameters to represent a desired target transmission rate, whereas we
  interpret them as a target amount of data enqueued at the network buffers.
  This latter interpretation has the added benefit of making the actual values
  of the parameters independent of the physical characteristics of the network
  elements. Although it can be argued that their interpretation is more to the
  letter of the original Vegas paper~\cite{brakmo94tcp}, our model is in
  greater accordance with actual
  implementations~\cite{ns-2,cardwell04:_tcp_vegas_implem_for_linux}, FAST
  description~\cite{wei06:_fast_tcp} and other Vegas
  models~\cite{samios03:_model_throug_of_tcp_vegas}.}

The vector $\vec a = \{0, a_1,a_2,\ldots,a_n\}$ can be calculated iteratively
by solving the equation $\sum_{i=1}^n x'_i/ C = 1$ starting with $n = 2$. While
there are simple algebraic formulae for this until $n$ reaches $4$, for greater
values only numerical procedures are possible, but a few iterations of the
Newton's method would suffice to give accurate results, for instance.

\begin{figure}
  \centering
  \resizebox{\columnwidth}{!}{
  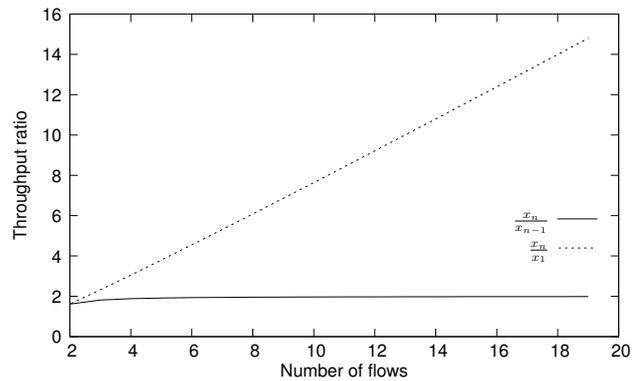}
  \caption{Throughput ratio achieved by the late coming flow relative to both
    the first flow and the previous one.}
  \label{fig:consecutive-effect}
\end{figure}
We have represented in Fig.~\ref{fig:consecutive-effect} the expected
throughput ratio of a newly arrived flow compared to both the oldest one and
the previous arrival for up to twenty consecutive flows. It can be seen how
even for a small number of consecutive arrivals the unfairness is severe. The
newest flow always gets twice the bandwidth of the previous arrival and the
oldest flow only gets one $n^\mathrm{th}$ of the newest flow throughput.

\subsubsection{Consequences in Buffer Dimensioning and Queueing Delay}
\label{sec:buffer-dimensioning}

The growth in the number of packets enqueued under persistent congestion can
have a dramatic effect on buffer dimensioning. Under ideal circumstances it
suffices to have a buffer size $\alpha$ times the number of possible FAST
flows to ensure that performance does not degrade. However, when taking into
account the extra packets enqueued under persistent congestion, the maximum
buffer size increases substantially.

\begin{figure}
  \centering
  \subfigure[Additional %
  packets per flow]{\includegraphics[width=0.45\columnwidth]{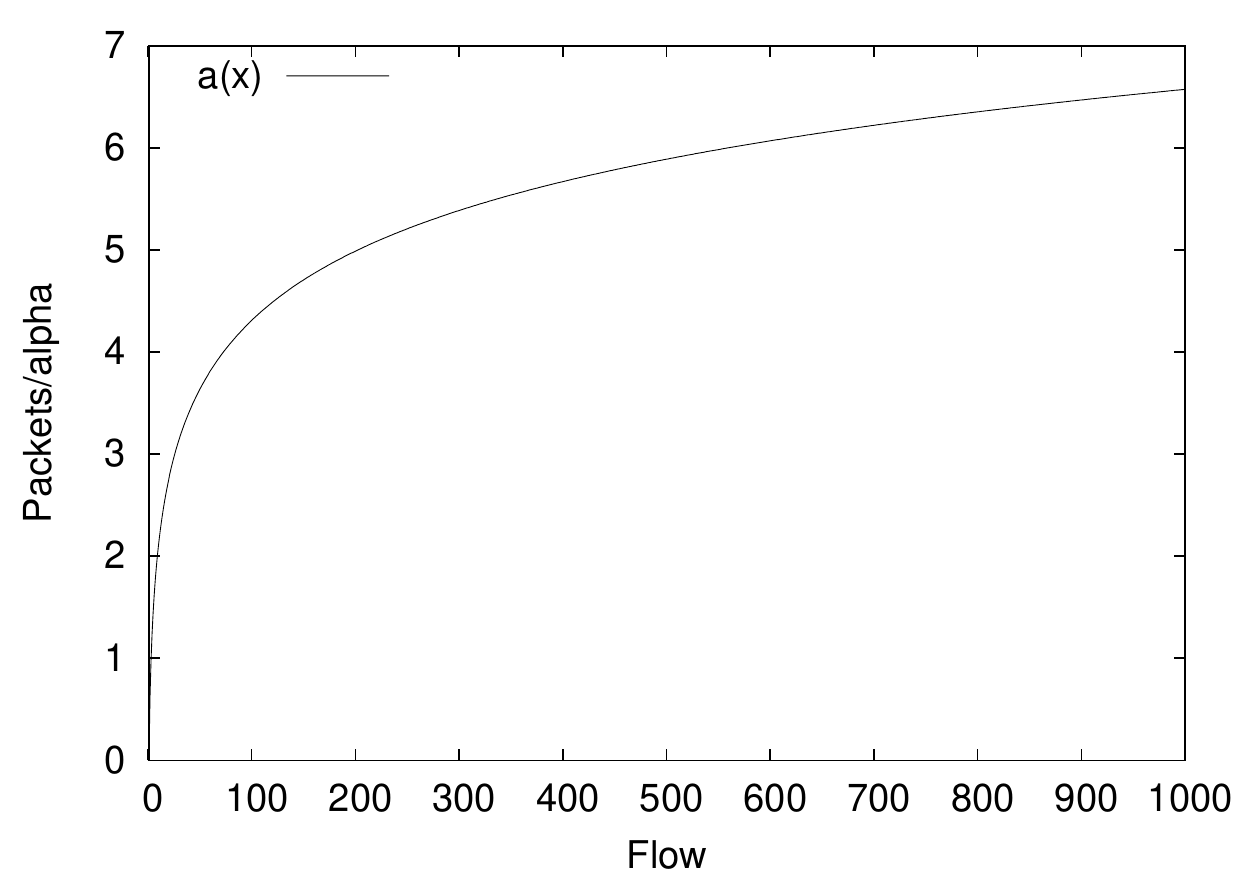}}
  \subfigure[Queue %
  length]{\includegraphics[width=0.45\columnwidth]{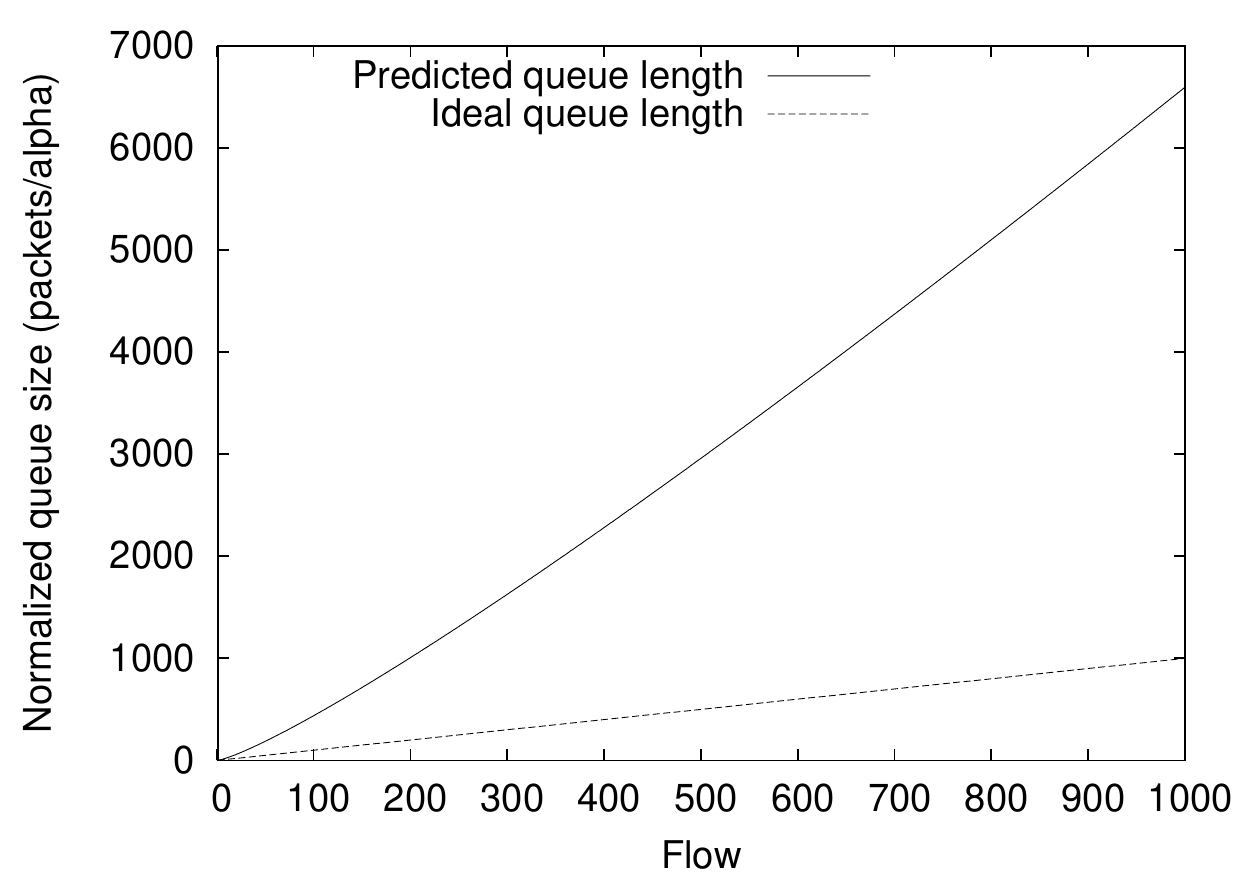}}
  \caption{Additional packets enqueued under persistent congestion and
    predicted queue length.}
  \label{fig:extra-packets}
\end{figure}
Fig.~\ref{fig:extra-packets} shows the values of vector $\vec a$ and the queue
length at the bottleneck router when up to $1\,000$ FAST flows are created
sequentially.\footnote{Although one thousand sequentially started flows is
  unlikely, it helps to show the asymptotic behavior of $\vec a$.} The number
of extra packets per flow needed to reach equilibrium ($a_i$) grows with the
logarithm of the number of flows. As a direct consequence, from around ten
flows onwards, each new flow enqueues more than $2 \alpha$ packets. In fact,
it is clearly seen that for one thousand flows the buffer size is almost seven
times larger than the calculated value for ideal behavior.

\subsection{One flow arriving to a $n$-flows stable scenario}
\label{sec:one-flow-arriving}

Another scenario of particular interest is the one in which a new flow arrives
when $n$ previous flows are fairly sharing the bottleneck bandwidth. This is
what would happen if the flows had a method to counter-measure the persistent
congestion problem, and thus we will employ this model in the following
sections to analyze the rate reduction approach presented in~\cite{cui06} and
to develop a new algorithm that reacts to persistent congestion in a
distributed fashion.

So, in our new scenario, there are $n$ flows with a correct estimation of the
propagation delay that, hence, have enqueued exactly $\alpha$ packets in the
network each. As a result, they all experience the same RTT of $\hat r^*_i =
n\cdot\alpha/C + \hat d^*_i, \forall i \leq n$. When the new flow arrives it
will estimate its propagation delay as $\hat d'_{n+1}=r_n=\ldots=r_1$, for a
target throughput equal to the bottleneck bandwidth $C$, leading to increased
queueing delay. When equilibrium is restored the flows measure $\hat r'_i =
\hat r^*_i+\frac{\alpha+\delta}{C}, i \le n$ and $\hat r'_{n+1}=\hat
d'_{n+1}+\frac{\alpha+\delta}{C}$. Defining $a$ again so as that
$\delta=\alpha\cdot a$, the throughput in equilibrium of all the flows can be
expressed as
\begin{eqnarray}
  \label{eq:tput_equil_nfast_old}
  & x'_i &= \frac{\alpha}{\frac{n\cdot\alpha+\alpha\cdot a}{C}} =
  = \frac{C}{n+1+a},~i\le n,\\
  \label{eq:tput_equil_nfast_new}
  & x'_{n+1} &
  = \frac{\alpha\cdot C}{\alpha\cdot(1+a)} = \frac{C}{1+a},
\end{eqnarray}
where $x'_i$ and $x'_{n+1}$ must hold
\begin{equation}
  \label{eq:tput-nstable-flows}
  \sum_{j=1}^n x'_i + x'_{n+1} = n\frac{C}{n+1+a} + \frac{C}{1+a} = C.
\end{equation}
Solving for $a$ in eq.~\eqref{eq:tput-nstable-flows} yields
\begin{equation}
  \label{eq:a-nstable}
  a = \frac{\sqrt{1 + 4 n}-1}{2}.
\end{equation}
We can see again how the unfairness in the bottleneck sharing is independent
of $C$ and $\alpha$, and only depends on the number of flows sharing the
resources. Moreover, this unfairness increases with the number of flows as
$\frac{x_{n+1}}{x_i} = O \left(\sqrt{n}\right), \forall i \leq n$.


\section{THE RATE REDUCTION APPROACH}
\label{sec:rate-reduction}

In this section we will analyze the rate reduction approach presented
in~\cite{cui06} to solve the persistent congestion problem. We will consider a
scenario similar to that analyzed in Section~\ref{sec:one-flow-arriving} in
which a late-coming flow arrives to a bottleneck being fairly shared by $n$
FAST flows. In a dynamic environment, when connections depart, the reduction
in throughput causes the occupancy of router buffers to drop, giving a chance
to remaining FAST connections to obtain a better estimate of their propagation
delays. Thus, it is reasonable to assume that the new connections must deal
with the presence of a number of existing flows aware of their true
propagation delays.

The proposed solution consists in restraining transiently the transmission
rate of a new flow by a given factor to allow router queues to get eventually
empty, thus giving new FAST connections a chance to measure the true
round-trip propagation delay.\footnote{The authors argue that the rate scaling
  factor should be set to a value similar to the $\alpha$~threshold.}
Unfortunately, and despite of the reduction on its rate, the new connection is
not always able to observe the empty queues. Note that, as the new flow drains
queues by reducing its own rate, competing flows respond by increasing their
rates. Hence, the new flow will only obtain the true propagation delay if
queues empty before existing flows are aware of this event, that is, if the
time required to empty the queues is less than the RTT of the existing flows.

The total backlog buffered at the core of the network in equilibrium, $B^*$, is
the sum of the backlog buffered by all active flows:
\begin{equation}
  B^* = \sum_{i=1}^{n+1} b^*_i = n\alpha + b^*_{n+1},
  \label{eq:backlog}
\end{equation}
where $b^*_i$ is the backlog buffered by flow $i$ in equilibrium. Assuming that
each flow $i, \forall i \leq n$, knows its true propagation delay, then these
connections will each maintain $\alpha$ packets in the router queues ($b^*_i =
\alpha$). On the other hand, the backlog buffered by the newly arrived flow
satisfies
\begin{equation}
   b^*_{n+1} = C (r^*_{n+1} - \hat d_{n+1}) = C \frac{\alpha}{x^*_{n+1}}\,.
   \label{eq:bn1}
 \end{equation}
Substituting \eqref{eq:tput_equil_nfast_new} and \eqref{eq:a-nstable} into
\eqref{eq:bn1}
\begin{equation}
  b^*_{n+1} = \alpha (1+a) = \frac{\alpha (1+\sqrt{1+4n})}{2}.
  \label{eq:bn2}
\end{equation}

This backlog will be drained from the queue at a rate equal to the bottleneck
link capacity minus the sum of the transmission rates of all active flows. In
the most favorable case, the new connection will completely pause its
transmission ($x_{n+1} = 0$). Considering the case in which all flows $i,
\forall i \leq n$, share a similar propagation delay ($d_i \approx d$) and
hence experience a similar RTT ($r^*_i \approx r^*$), the fairness condition
becomes
\begin{equation}
  \frac{B^*}{C - \sum^{n}_{i=1} x^*_i} < r^* = d + \frac{B^*}{C}.
  \label{eq:faircond1}
\end{equation}
Finally, substituting \eqref{eq:tput_equil_nfast_old}, \eqref{eq:backlog} and
\eqref{eq:bn2} into eq.~\eqref{eq:faircond1}, it follows that
\begin{equation}
  d > \frac{n \alpha \left( 1+\sqrt{1+4n} \right)}{2 C} =
    \frac{n b^*_{n+1}}{C}.
  \label{eq:faircond2}
\end{equation}
Thus, the rate reduction method is only effective when the round-trip
propagation delay of competing flows exceeds the lower bound calculated
in~\eqref{eq:faircond2}. Note that the lower bound scales as
$O\left(n^{3/2}\right)$ with the number of active flows.


\section{OUR SOLUTION}
\label{sec:our-prop}

In this section we will present a solution to the persistent congestion
problem that lacks the rate reduction method limitations. We noticed that,
when the newly arriving flow stabilizes, it can indirectly obtain a good
estimation of its actual round-trip propagation delay.

As we pointed in Section~\ref{sec:one-flow-arriving}, the new flow
overestimates its propagation delay as $\hat d_{n+1} = d_{n+1} + \sum_l
t_{\mathrm{tx}_l} + n \alpha/C$. That is, there is an error $e = n\,\alpha/C$.
Therefore, provided the new flow knows (or accurately guesses)~$n$ and the
bottleneck link capacity, $C$,a more precise estimate of the propagation delay
could be calculated as
\begin{equation}
 \hat d'_{n+1} = \hat d_{n+1} - \hat e,~\text{ with } \hat e= \hat n
 \frac{\alpha}{\hat C},
\end{equation}
where $\hat{n}$ and $\hat{C}$ are the inferred values of $n$ and $C$,
respectively.

In order to obtain good values for $\hat n$ and $\hat C$ it suffices to induce
short variations in the throughput of the late coming flow and measure the
changes it produces in queueing delay. Let $r^*_{n+1}$ be the RTT of
the newest flow once it reaches a stable throughput. If this connection
modifies its transmission rate, for instance by changing the value of the
window size, $w^{\epsilon}_{n+1} = (1-\theta)w^*_{n+1}$ with $\theta < 1$ for a brief
time $t_{\epsilon}$ it will measure a new round trip time $r_{n+1}^{\epsilon}$
after this time.\footnote{Note that using positive values for $\Theta$ can
  cause the queues to deplete before the time $t_{\epsilon}$ is over, thus
  rendering the following analysis inaccurate. This can be avoided using small
negative values for $\Theta$ causing the queueing delay to increase. Although
this can lead to packet drops in insufficiently dimensioned routers, this
situation is easily detected and avoided using smaller values for $\Theta$ in
posterior measures.} Let $\Delta r_{n+1} = r_{n+1}^{\epsilon} - r^*_{n+1}$.
Under such circumstances
\begin{equation}
  \label{eq:depletion}
  C\Delta r_{n+1} = \left(
    C-\sum_{i=1}^n \frac{w^*_i}{r^*_i} - (1-\theta)\frac{w^*_{n+1}}{r^*_{n+1}}
    \right) t_{\epsilon},
\end{equation}
as long as $t_{\epsilon}$ is short enough so that the transmission rate of the
first $n$ sources remains constant. For this it is enough to make
$t_{\epsilon}$ of the same order as $r^*_i$.

A estimation of $\hat n$ can be directly obtained
substituting~\eqref{eq:tput_equil_nfast_old},~\eqref{eq:tput_equil_nfast_new}
and~\eqref{eq:a-nstable} into~\eqref{eq:depletion} and using the fact that
$x=\frac{w}{r}$. Solving for $\hat n$,
we reach
\begin{equation}
  \label{eq:n-estimation}
  \hat n = \frac{\theta t_{\epsilon}}{\Delta r_{n+1}} \left(
    \frac{\theta t_{\epsilon}}{\Delta r_{n+1}} - 1
    \right).
\end{equation}
Once we have $\hat n$ it is trivial to obtain $\hat C$
using~\eqref{eq:tput_equil_nfast_new} and~\eqref{eq:a-nstable}, obtaining
\begin{equation}
  \label{eq:c-estimation}
  \hat C = \frac{\left(
      1+\sqrt{1+4n}
    \right)w_{n+1}}{2 r_{n+1}}.
\end{equation}

The proposed adjustment of the round-trip propagation delay suffices to solve
the persistent congestion problem since transmission rates of competing flows
will eventually converge to their expected values. As the former method, our
proposal can be applied without the need of any network support but, in this
case, the propagation delay of competing flows does not affect their behavior.

Additionally, the estimates obtained with our proposal can be used to reduce
convergence time by computing an optimal value for the congestion window of
the newly arrived flow. Recall that, when these estimates are obtained, the
new flow is enjoying more bandwidth than its fair share, so it should reduce
its congestion window to a more suitable value as well. Ideally, all the flows
must maintain $\alpha$ packets buffered into the core of the network. In such
circumstances, the RTT experienced by the new flow should be $r^{\prime}_{n+1}
= d^{\prime}_{n+1} + (n+1)\alpha/C$. Therefore, since $w = r x$, the value
that should be assigned to the congestion window is
\begin{equation}
  w^{\prime}_{n+1} = \frac{\alpha r^{\prime}_{n+1}}{r^{\prime}_{n+1}-d^{\prime}_{n+1}}
  = \alpha + \frac{d^{\prime}_{n+1} \hat{C}}{\hat{n} + 1}.
  \label{eq:wideal}
\end{equation}


\section{EXPERIMENTAL VALIDATION}
\label{sec:exper-valid}

In this section we present the results of two series of experiments. The first
group tests the validity of the model of the persistent congestion problem
presented in Section~\ref{sec:analytical-model}. The second group verifies the
appropriateness of our solution to the persistent congestion problem,
available for download at~\cite{herreria09:_fast_tcp_conges_avoid_implem},
when compared with the rate reduction approach~\cite{cui06} and with the
original FAST protocol.

\begin{figure}
  \centering
  \includegraphics[width=\columnwidth]{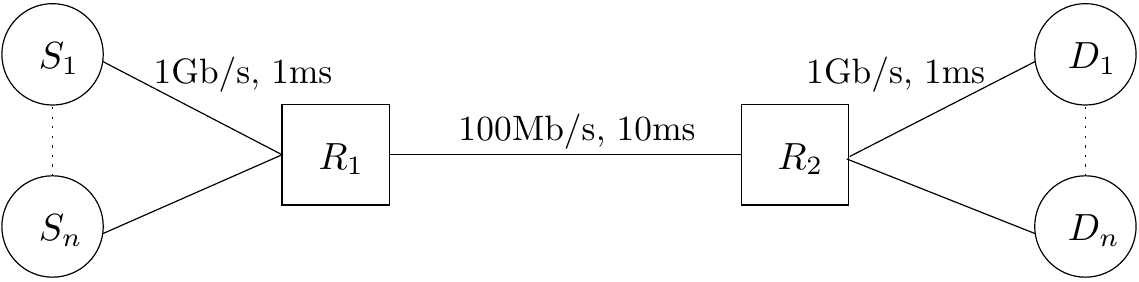}
  \caption{Network topology used in the simulation experiments.}
  \label{fig:network-model}
\end{figure}

All the experiments were simulated with version 2.31 of ns-2~\cite{ns-2} and,
unless otherwise noted are based on the scenario depicted in
Fig.~\ref{fig:network-model} with little variations explained in each
experiment.
We establish a FAST-TCP connection between each source $S_i$ and its
corresponding destination $D_i$. Unless otherwise noted, each connection is
configured with $\alpha=50$ packets and all packets carry a payload of
$1\,000$ bytes. Because in this paper we are only concerned with the
congestion avoidance characteristics of FAST, buffer sizes are big enough to
hold all the packets enqueued by the flows, as predicted by
eq.~\eqref{eq:n-flows-rtt}.

\subsection{Analytical Model}
\label{sec:analytical-model-val}

The following experiments are all designed to validate our modeling of the
persistent congestion problem.

\subsubsection{Two flows scenario}
\label{sec:exp-two-flows-scenario}

We start with the simple two-flows scenario and validate the claims presented
in Section~\ref{sec:two-flows}, namely that the unfairness is independent on
both FAST configuration and network characteristics. To this end we simulate
the previously described test network both under ideal conditions, i.e., with
no background traffic, and in a noisy environment.

\begin{figure}
  \centering
  \includegraphics[width=\columnwidth]{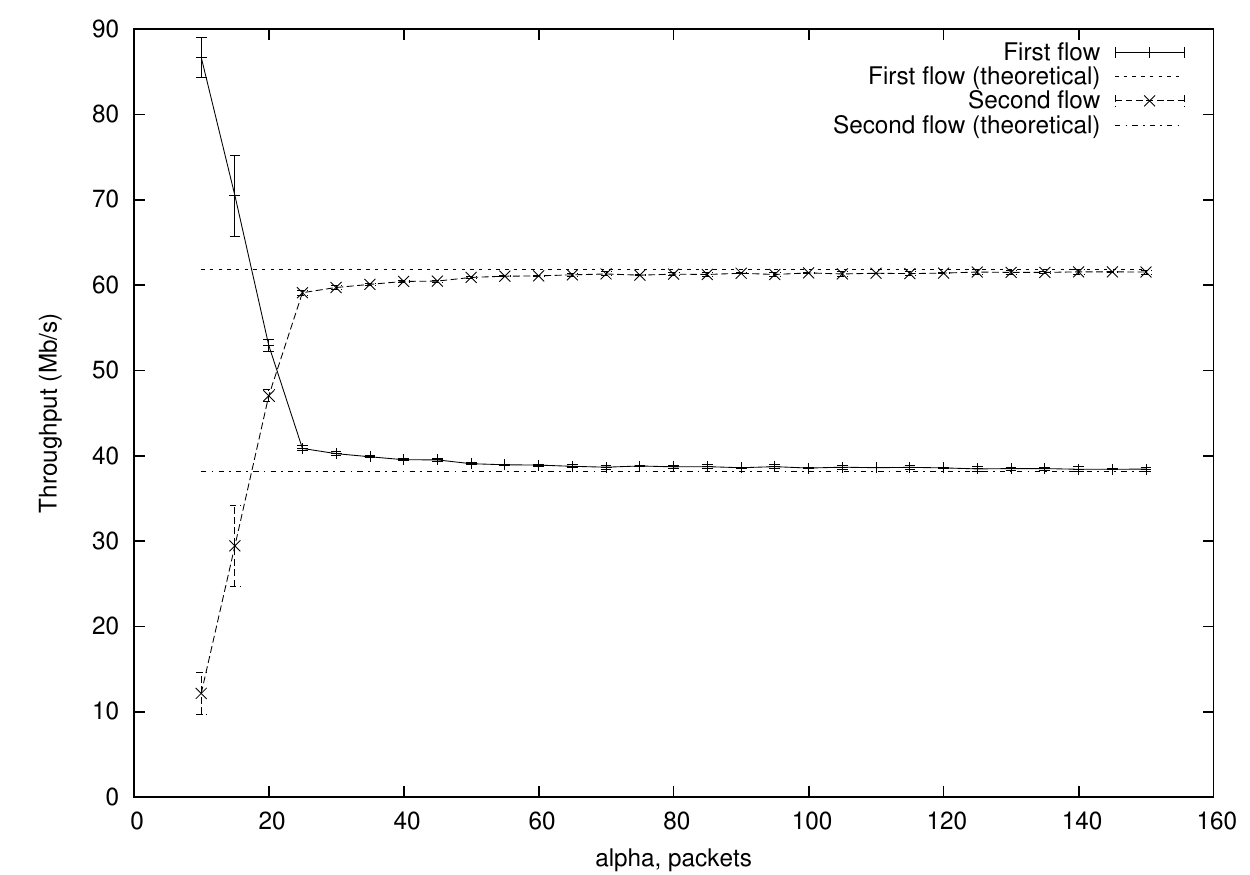}
  \caption{Average throughput obtained by both flows for different values of
    $\alpha$. Error bars show a 95$\,\%$ confidence interval.}
  \label{fig:varbeta}
\end{figure}
Fig.~\ref{fig:varbeta} plots the throughput obtained by two FAST flows started
sequentially when we vary the value of $\alpha$.\footnote{In this and
  following figures, error bars correspond with a 95\% confidence interval.}
Flows were started with a 2 seconds gap to ensure the first flow have had
plenty of time to achieve its steady state and an uniformly distributed random
time (between 0 and 1 second) to add some randomness. In this experiment we
have run both flows for $10\,\mathrm{s}$ and plotted the averaged throughput
of both flows after $25$ simulations with slightly different starting times.
It can clearly be seen that the theoretical results hold for almost any value
of $\alpha$. The deviations when $\alpha<30$ are caused by inherent
instabilities in FAST when $\alpha$ is too small for the network. With such
small $\alpha$ values FAST is unable to converge. The stability
characteristics of FAST have been extensively studied in the literature, in
fact~\cite{wang05:_model_and_stabil_of_fast_tcp,choi05:_global_stabil_of_fast_tcp,choi06:_global_expon_stabil_of_fast_tcp}
give sufficient conditions $\alpha$ must meet to avoid this problem.

In order to test how the modeling behaves in more stringent scenarios, with
non-100\% FAST traffic, we have repeated the above experiment adding some
background traffic to the bottleneck link. This background traffic was
simulated with several Pareto traffic sources on top of UDP in a similar
fashion as in~\cite{wei06:_fast_tcp}. Each Pareto flow was set up with a shape
factor of $1.5$, average burst and idle time of $100\,$ms and a peak rate of
$1\,$Mb$/$s, thus consuming, on average, $0.5\,\%$ of the bottleneck
bandwidth.

\begin{figure}
  \centering
  \includegraphics[width=\columnwidth]{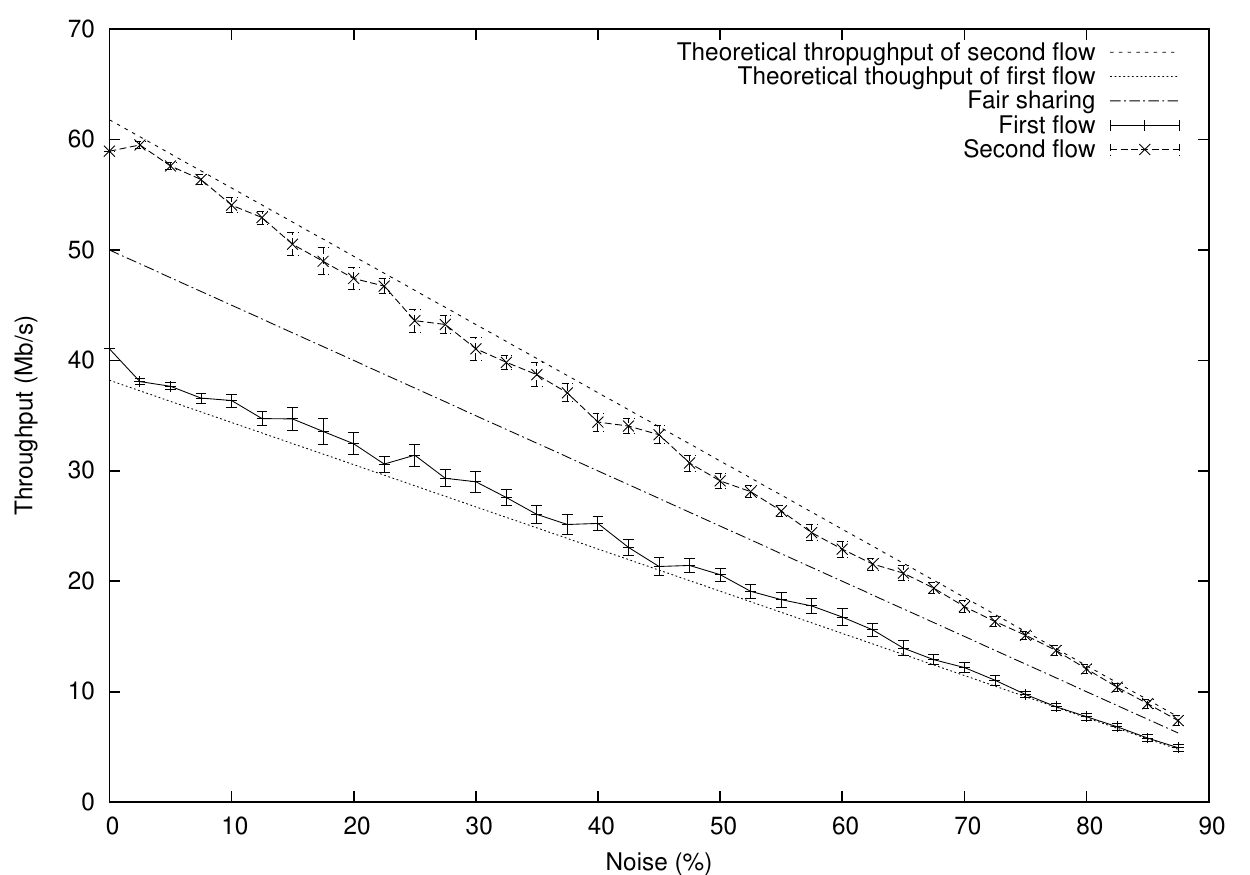}
  \caption{Average throughput of two sequentially started FAST flows in the
    presence of background traffic.}
  \label{fig:noise}
\end{figure}

We have run simulations for a different number of background flows (from none
to two hundred flows) and represented the results in Fig.~\ref{fig:noise}.
It can be observed how, despite the high amount of noise, that reaches the
full bandwidth of the bottleneck link, results match those predicted. That is,
both flows share in an unfair manner the bandwidth not used by the noise.



\subsubsection{FAST flows arriving sequentially}
\label{sec:exp-sever-flows-arriv}

This second set of experiments measures the impact of persistent congestion in
a worst case scenario: flows arriving sequentially at a bottleneck link.

The first experiment compares the relative throughput obtained by each
sequentially started flow under an all-FAST scenario and for a different
number of total flows, from just two flows up to nine.

\begin{figure}
  \centering
  \includegraphics[width=\columnwidth]{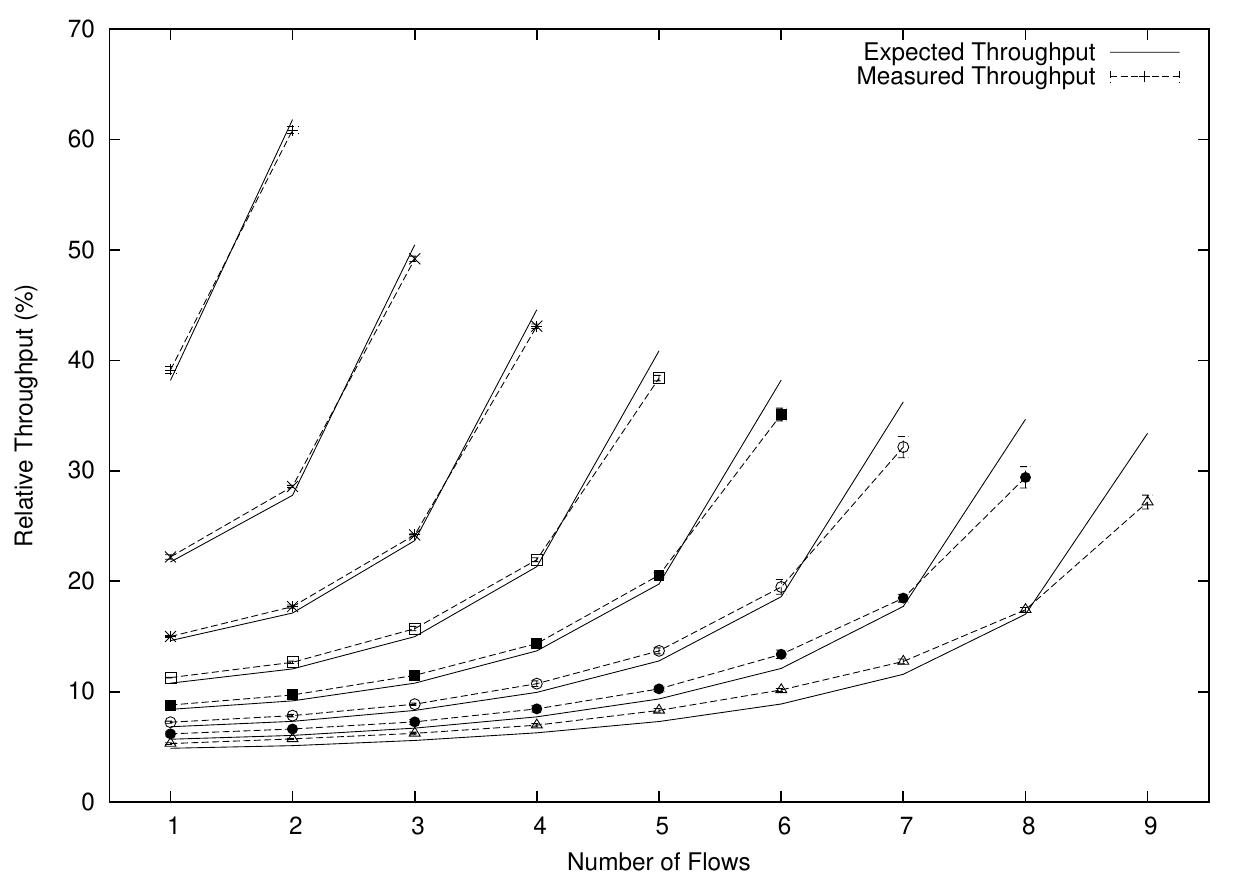}
  \caption{Measured throughput for different number of flows started
    sequentially. Each line joins the final relative throughput of each flow
    for each experiment (same number of total flows).}
  \label{fig:tput}
\end{figure}

Fig.~\ref{fig:tput} shows both the predicted and the measured throughput. 
The measured throughput is the result of averaging the values obtained for
different values of $\alpha$ (between $40$ and $60$). For easier observation,
the results corresponding to the same number of total flows are joined by a
continuous line. That is, there are 8 lines, one for the two flows experiment,
a second one joining the throughput of the flows in the three flows
experiment, and so on. Each line has as many points as flows, each point
representing the averaged throughput obtained by the $i$-th coming flow. The
model produces very accurate predictions that match the values obtained by
simulation. These results agree with those observed in other works where the
asymmetry in throughput among Vegas flows was first pointed
out~\cite{hengartner00:_tcp_vegas_revis,mo99:_analy_compar_tcp_reno_vegas}.

\begin{figure}
  \centering
  \includegraphics[width=\columnwidth]{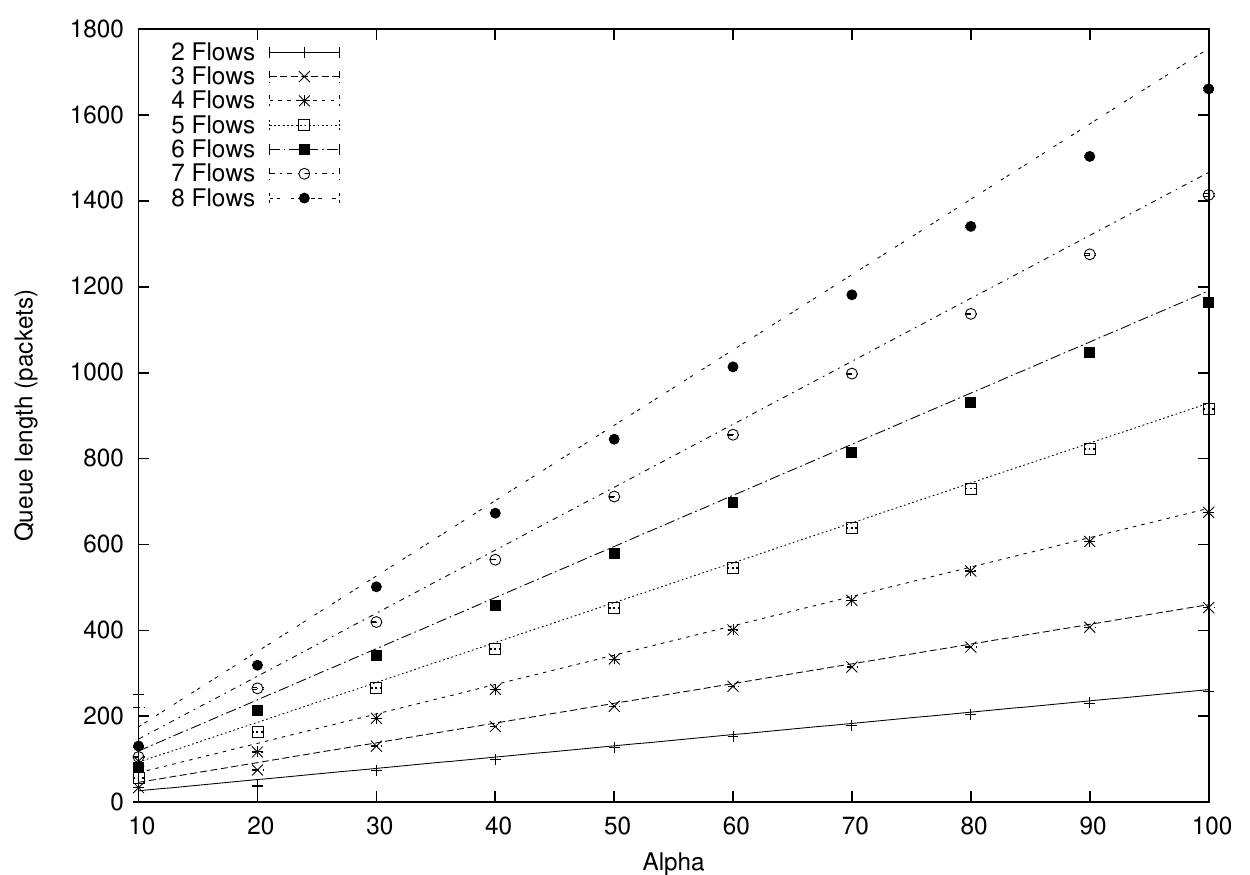}
  \caption{Measured queue length for different number of flows started
    sequentially. Continuous lines show the values as predicted by the model
    in Section~\ref{sec:buffer-dimensioning}.}
  \label{fig:queue-length}
\end{figure}

Persistent congestion does not only have adverse effects on fairness, but
transmission delay worsens as well, as buffer occupancy grows larger that
expected. In fact, the results in Fig.~\ref{fig:queue-length} confirm that
queue sizes grow much larger than $\alpha$ times the number of flows.
These results clearly show that the model predictions are quite accurate. The
small differences in the plot are somewhat misleading: they are due to the
fact that the model computes real values for the queue length, while the
length obtained in the simulations comes expressed in integer units.

\subsubsection{One flow arriving to a $n$ flows stable scenario}
\label{sec:exp-one-flow-arriving-1}

Finally we test the model for the scenario we are most interested in. This is
the situation that happens in a network shared by FAST flows enhanced with
some mechanism to correct persistent congestion. The experiment thus test the
validity of eq.~\eqref{eq:a-nstable}, that forms the basis for our solution to
the persistent congestion problem presented in Section~\ref{sec:our-prop}.

\begin{figure}
  \centering
  \includegraphics[width=\columnwidth]{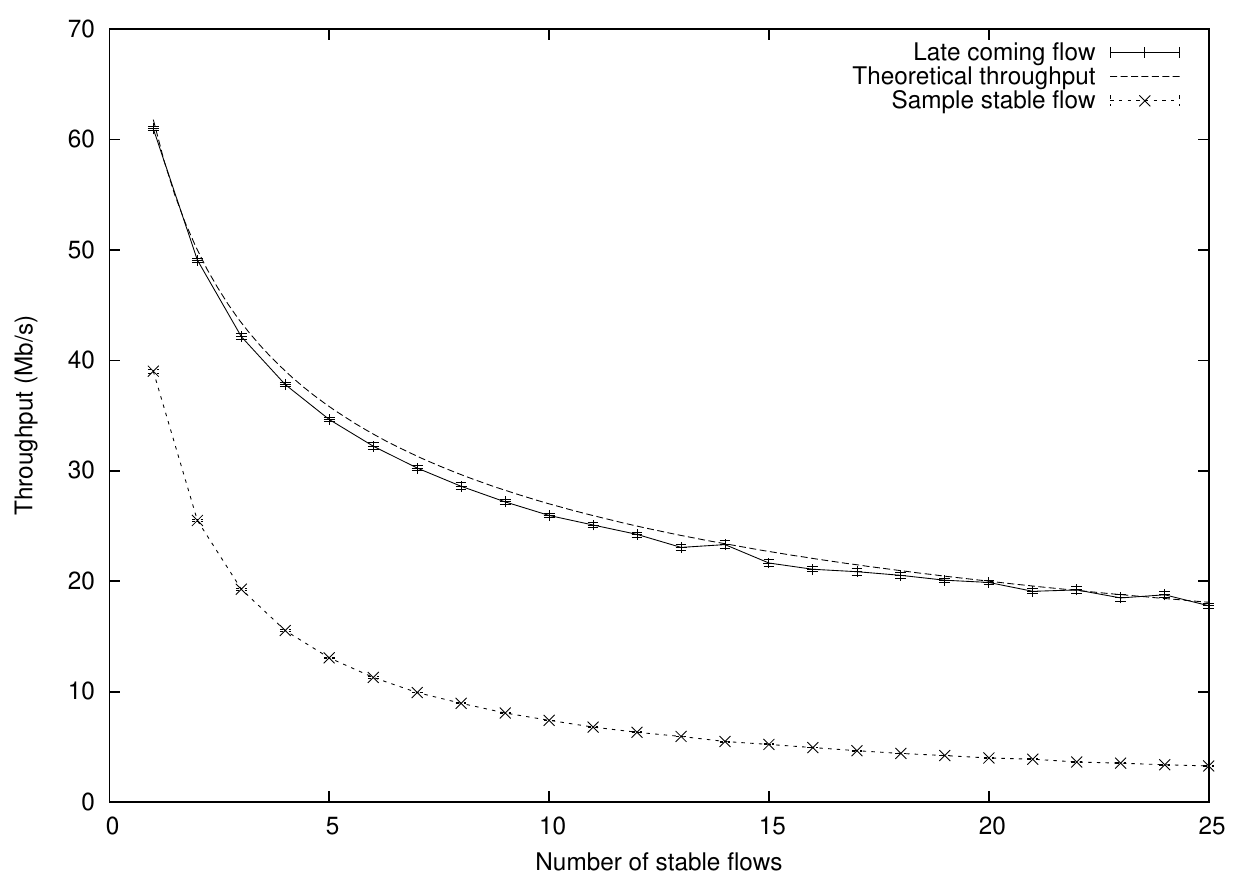}
  \caption{Throughput of a late coming flow when confronted with several FAST
    flows already getting their fair share.}
  \label{fig:varnstable}
\end{figure}

In order to obtain $n$ flows with the proper estimation propagation delay
without using yet our modification we let each one to run for some time in
isolation to later restart the $n$ flows simultaneously. After all $n$ flows
reach equilibrium, we start the late coming flow and measure the throughput of
the latter and a representative flow from the initial set. We have employed a
fixed packet size of $1\,000$ bytes and different values of $\alpha$ for each
simulation. The results, for different values of $n$, are plotted in
Fig.~\ref{fig:varnstable}.

\subsection{Solutions}

We have also conducted several simulation experiments to verify our claims
regarding the rate reduction approach and validate our proposal. We have
implemented both methods in the ns-2 simulator. We have employed the same
network topology (Fig.~\ref{fig:network-model}) and configuration parameters
than in the previous experiments.
 
\subsubsection{Impact of the value of $\theta$}
\label{sec:impact-value-theta}

A precise estimation of the number of flows is essential to remove the error
introduced by persistent congestion on the measured propagation delay. The
following experiment measures the accurateness of this estimation for
different values of $\theta$.

The simulation is as follows. A set of FAST flows aware of their true
propagation delay share a single bottleneck link. Once their (equal) rates
stabilize, a new flow using our proposed measurement method starts. The
simulation is repeated ten times varying slightly the starting time of the
flows.
\begin{figure}
  \centering
  \includegraphics[width=\columnwidth]{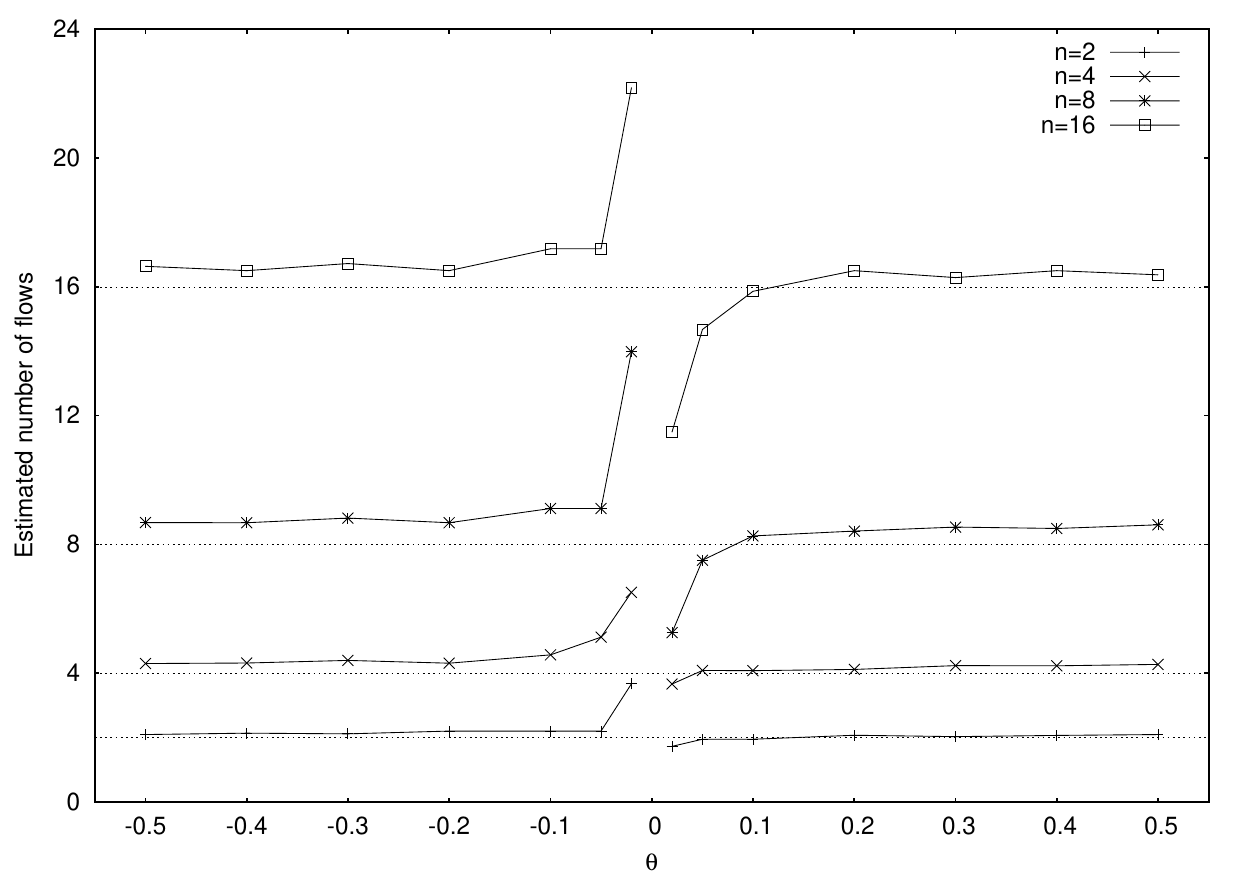}
  \caption{Impact of $\theta$ on the estimation of the number of flows ($n$).}
  \label{fig:thetavsn}
\end{figure}
Fig.~\ref{fig:thetavsn} shows the number of existing flows estimated by the
late coming flow using different values of $\theta$ for different values of
$n$, the number of existing flows.\footnote{Although $95\,$\% confidence
  intervals have been calculated, they are not represented since they were
  consistently lower than $\pm 1\,$\% and just cluttered the figure.} The
proposed method is unable to obtain reliable estimates when $\theta$ gets too
close to~$0$. However, variations from $|\theta| = 0.1$ onwards avoid
undesired deviations. In the following experiments, we have employed
$\theta=-0.5$ to acquire good estimates while preventing the bottleneck from
getting empty at the same time.

\subsubsection{Effectiveness of our proposal}

Firstly, we have simulated a scenario where five FAST connections are sharing
the bottleneck link. The sources are started at intervals of $20\,$s
each.\footnote{Only consecutive arrivals were considered. Departures are a
  trivial case if we assume that the system converges to a fair share. When a
  flow leaves the network, the queue occupation eventually just diminishes in
  $\alpha$ packets, and the situation is no different than that of $n-$ flows
  already sharing fairly a bottleneck link, with $n$ being the number or
  previous flows.}
\begin{figure}
  \centering
  \subfigure[FAST TCP]{\includegraphics[width=0.45\columnwidth]{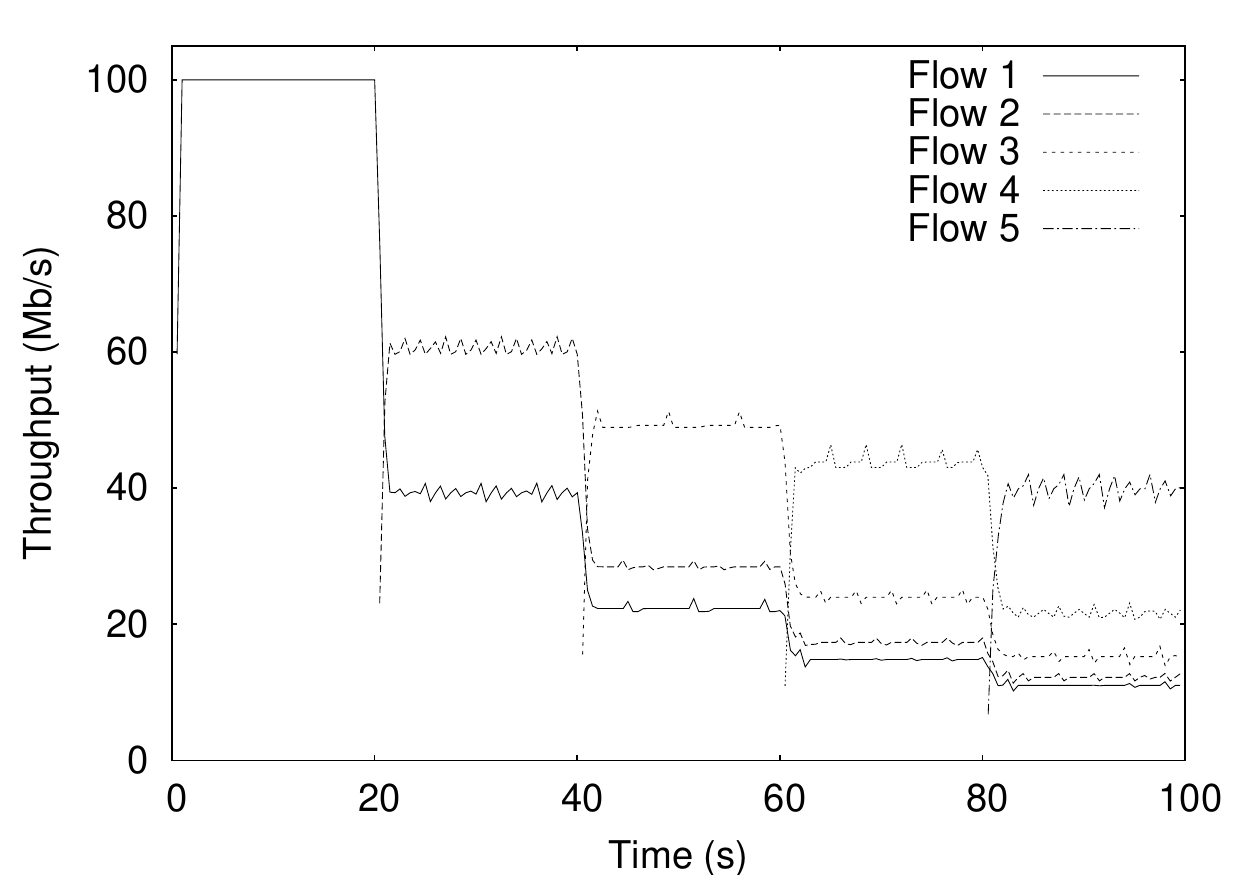}
    \label{fig:tputfast}}
  \subfigure[Modified FAST]{\includegraphics[width=0.45\columnwidth]{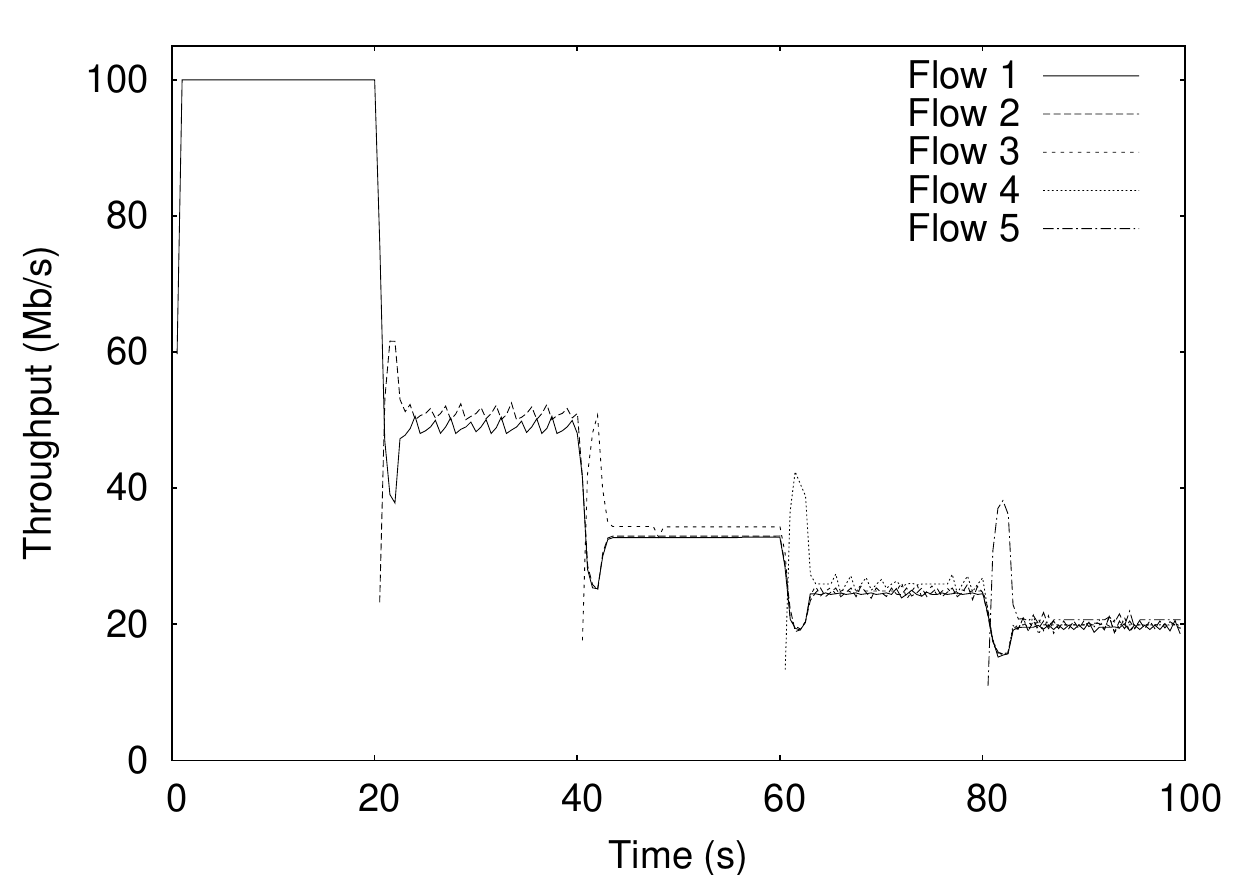}
    \label{fig:tputenh}}
  \caption{Throughput comparison.}
  \label{fig:tput-comparison}
\end{figure}
Fig.~\ref{fig:tputfast} shows the instantaneous throughputs of the FAST
connections when the original congestion avoidance mechanism is used.
As expected, FAST strongly favors new sources and recent
connections enjoy larger throughputs compared to old connections. When our
method is applied, this bias disappears and the network bandwidth is shared
among competing FAST connections in a fair manner (Fig.~\ref{fig:tputenh}).

\begin{figure}
  \centering
  \includegraphics[width=\columnwidth]{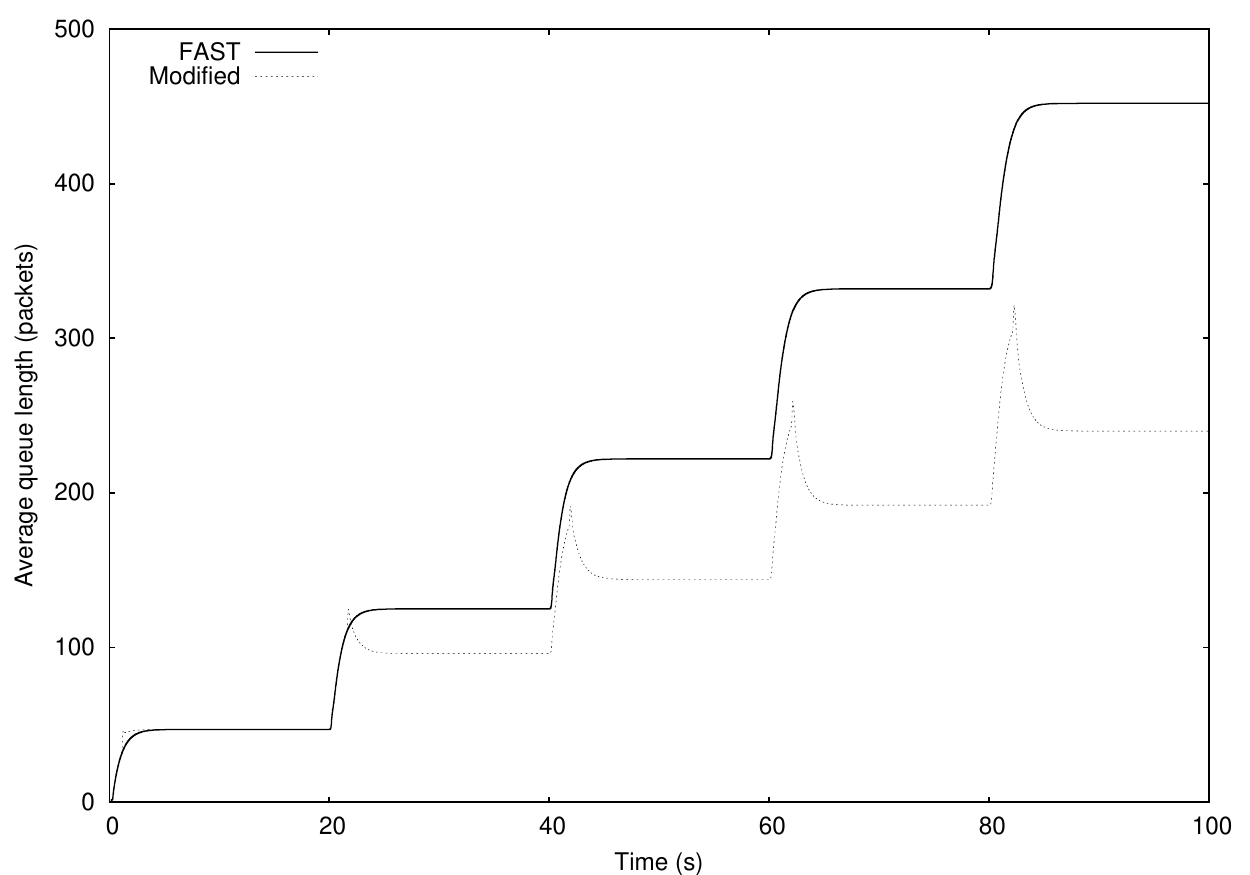}
  \caption{Core queue length comparison.}
  \label{fig:cqueue-comparison}
\end{figure}
The average queue length at the bottleneck is also shown in
Fig.~\ref{fig:cqueue-comparison}. Due to persistent congestion, the amount of
extra data introduced by FAST is larger than the targeted amount ($\alpha$
packets per connection). However, our proposal keeps the proper amount of
extra data into the network and thus the average queue size is smaller.

\subsubsection{Impact of different round-trip propagation delays}

In the second experiment, we have examined the impact of different round-trip
propagation delays on bandwidth distribution. We consider a number of existing
FAST flows knowing their true propagation delays and, therefore, sharing the
available bandwidth uniformly. Once their transmission rates have stabilized,
a new flow starts its transmission. In order to study the effect of different
propagation delays, the delay of the link between nodes $R_1$ and $R_2$ has
been changed from $3$ to $53\,$ms. To evaluate the fairness among the new and
the existing connections, we used the following ratio:
\begin{equation}
  \mathit{Fairness~Ratio} = \frac{n \bar{x}_{n+1}}{\sum^{n}_{i=1} \bar{x}_i}\,,
\end{equation}
where $n$ is the number of existing flows, $\bar{x}_{n+1}$ is the average
transmission rate of the new flow and $\bar{x}_i$ is the average transmission
rate of existing flow $i=1,\ldots,n$. Clearly, if the new connection obtains
the same throughput as its competitors, the ratio will be $1$.
\begin{figure}
  \centering
  \subfigure[$n=4$]{\includegraphics[width=0.45\columnwidth]{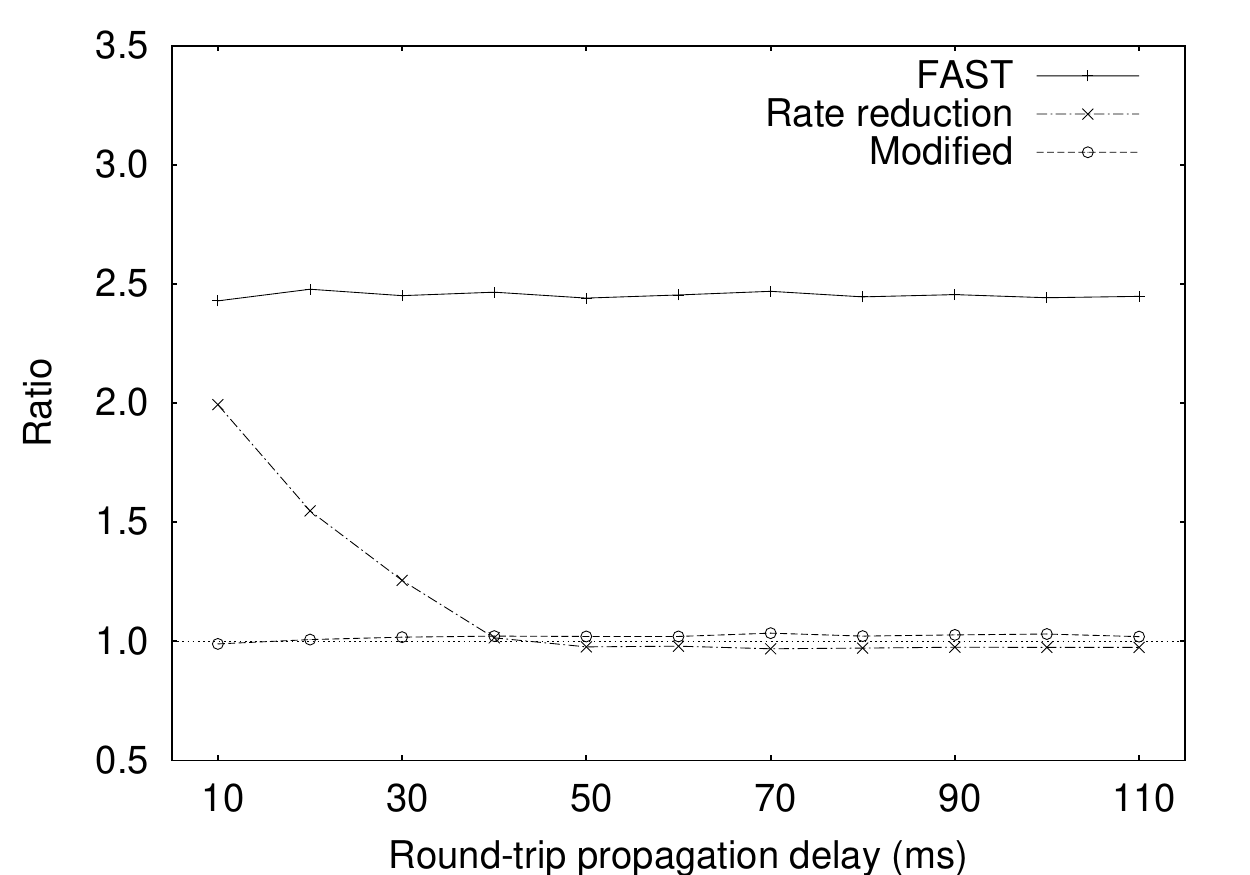}
    \label{fig:ratio4}}
  \subfigure[$n=8$]{\includegraphics[width=0.45\columnwidth]{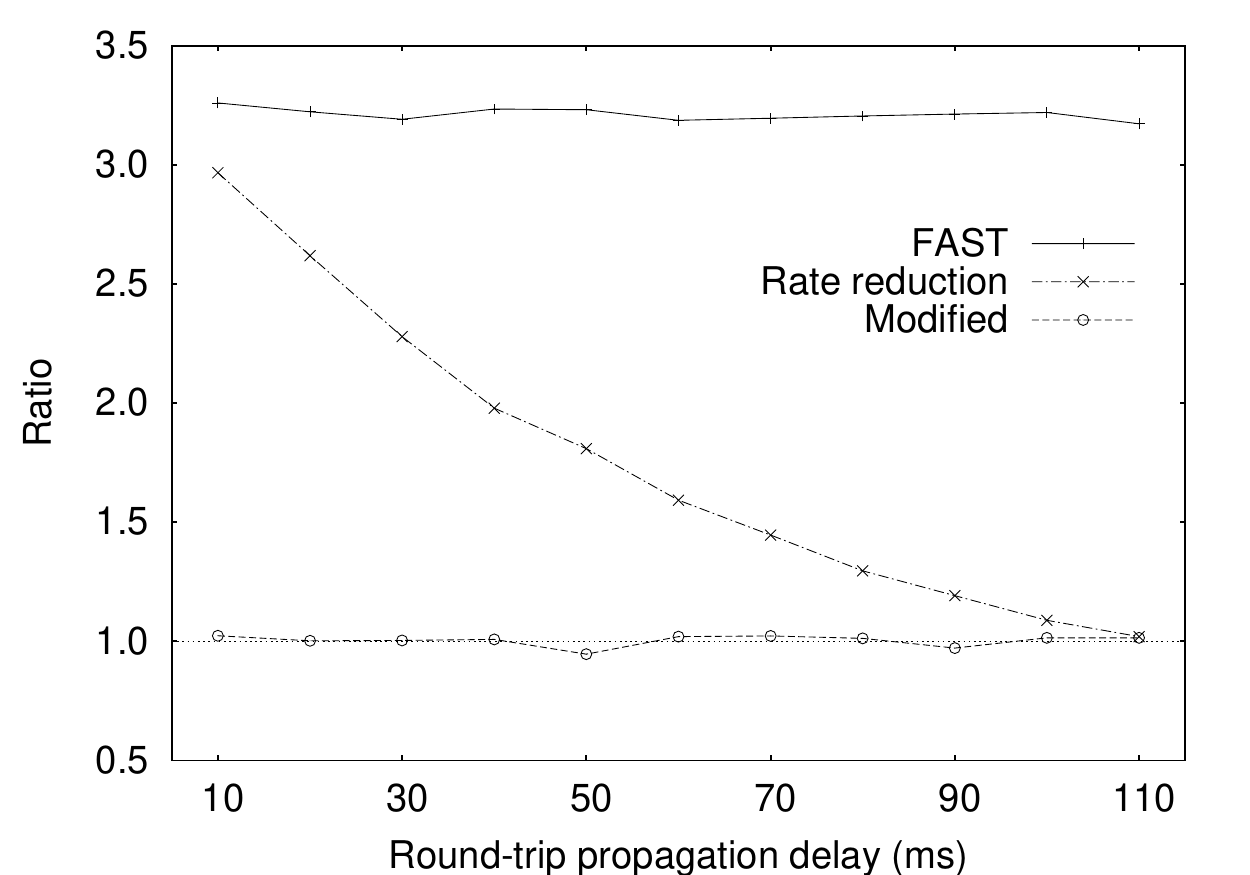}
    \label{fig:ratio8}}
  \caption{Impact of round trip propagation delay.}
  \label{fig:ratio}
\end{figure}
Fig.~\ref{fig:ratio} compares the performance of original FAST-TCP, the rate
reduction approach and our proposal for two different values of~$n$. As
expected, with FAST-TCP, the new connection obtains a higher throughput. With
the rate reduction method, the bandwidth sharing depends on the experienced
propagation delay. The minimum round-trip propagation delays required for the
rate reduction approach to work properly as calculated using
eq.~\eqref{eq:faircond2} are $40.9\,$ms for $n=4$ and $107.9\,$ms for $n=8$.
Graphs show how, as the propagation delay of existing flows falls below these
thresholds, the bandwidth distribution becomes less fair. In contrast, with
our proposal, fairness is preserved in all simulated scenarios.

\subsubsection{Impact of the number of flows}
\label{sec:impact-number-flows}

We have also compared our proposal to the rate reduction approach when the
number of flow increases, while maintaining the rest of the simulation
parameters fixed.

\begin{figure}
  \centering
  \includegraphics[width=\columnwidth]{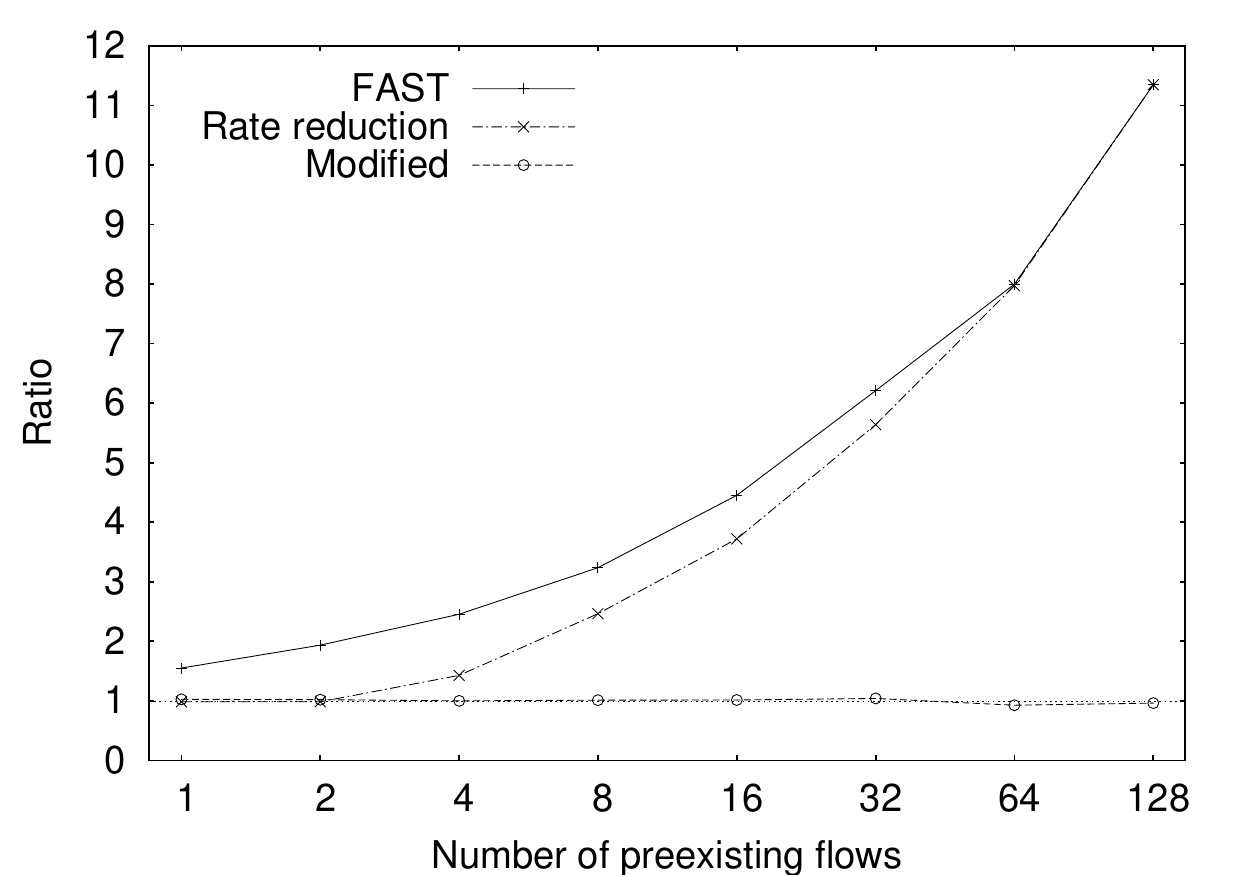}
  \caption{Impact of the number of preexisting flows.}
  \label{fig:ration}
\end{figure}

Fig.~\ref{fig:ration} shows that, while our solution manages to stay fair
irrespectively of the number of flows, the rate reduction approach deviates
from fairness and approximates original FAST behavior as the number of flows
increases.

\subsubsection{Impact of background traffic}
\label{sec:impact-backgr-traff}

We end the validation section presenting a non-ideal scenario. For this we use
a more realistic and stringent topology and noisy background traffic that
interferes with our estimation method.
\begin{figure}
  \centering
  \includegraphics[width=\columnwidth]{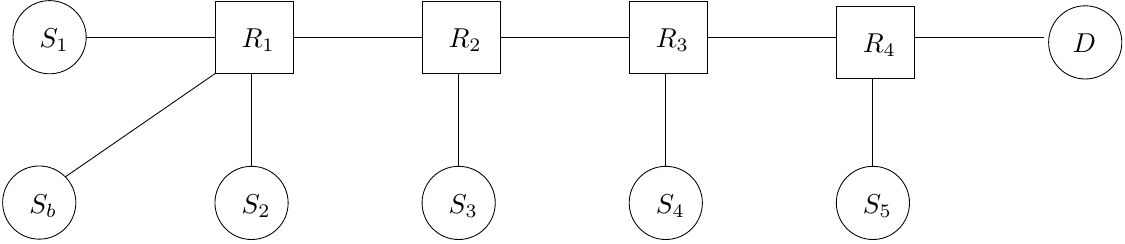}
  \caption{Parking lot topology with multiple bottlenecks. Every link has a
    $100\,$Mb$/$s capacity and a $5\,$ms propagation delay.}
  \label{fig:mnet}
\end{figure}
Fig.~\ref{fig:mnet} shows the topology employed, a variant of the classic
parking-lot topology. The various bottlenecks are traversed by five flows
running from nodes $S_1, ..., S_5$ towards node $D$. The flow from node $S_1$
starts its transmission last, giving time to the previous flows to stabilize.
Additionally, in a similar way as in~\cite{wei06:_fast_tcp}, some background
traffic was simulated via a Pareto flow $(S_b,D)$ with a shape factor of
$1.25$, average burst and idle time of $100\,$ms and a peak rate ranging from
$5$ to $100\,$Mb$/$s.
\begin{figure}
  \centering
  \includegraphics[width=\columnwidth]{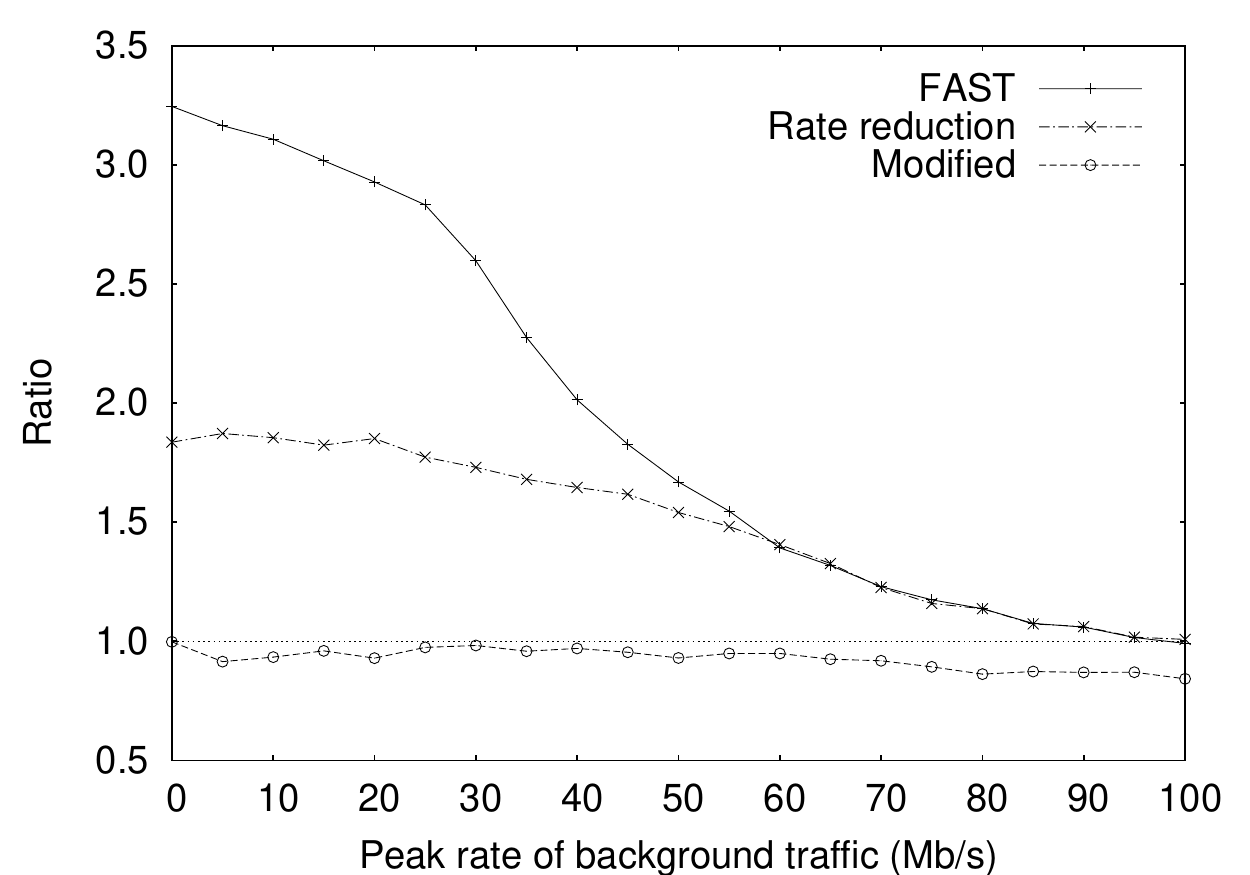}
  \caption{Impact of background traffic.}
  \label{fig:bgtraffic}
\end{figure}
Fig.~\ref{fig:bgtraffic} shows the results. Not surprisingly, with both FAST
and the rate reduction method, fairness improves as the peak rate of
background traffic increases. The reason is that during active periods FAST
flows reduce their rates as router queues fill up due to background traffic.
When the background noise diminishes (idle periods) queues drain as the total
FAST traffic is smaller than the link capacity and flows can seize a better
estimate of their respective propagation delays before queues start to fill up
again. In contrast, our solution deviates from absolute fairness when the
background noise gets too high, because it interferes with our estimation
method. However, it reaches a fairness index of $0.84$ even with a peak noise
level equal to the bottleneck bandwidth and obtains significantly better
results than both original FAST and the Rate reduction method when there is
less background traffic. This is a very good result if we keep in mind that
our method was designed for the case when there is just FAST traffic in the
network.


\section{CONCLUSIONS}
\label{sec:conclusions}

Taking as a starting point the FAST model in~\cite{wei06:_fast_tcp} we have
established explicit formulæ that predict the throughput of a given DCA flow
under persistent congestion conditions. We have found that assuming all flows
have the same configuration parameters, the bandwidth share is independent of
both their actual values and network configuration.

We have employed the aforementioned model to analyze \emph{the rate reduction
  approach,} one of the most promising end-to-end proposals that try to deal
with the persistent congestion problem. We have found that it does not work in
every network configuration. In fact, we have provided necessary conditions a
network must meet for the rate reduction approach to be useful.

Finally, we have presented an amendment to FAST that makes it immune to the
persistent congestion problem. We used our analysis to give FAST senders the
ability to discern persistent congestion and react accordingly. Our proposed
solution outperforms previous approaches and does not need network
modification to work. In fact, it is insensible to the values of propagation
delay and bottleneck capacity.


\acks This work was supported by the ``Ministerio de Educación y Ciencia''
through the project TIC2006-12507-C03-02 of the ``Plan Nacional de I+D+I''
(partially financed with FEDER funds).

\bibliographystyle{wileyj}
\bibliography{IEEEfull,biblio}

\begin{thebibliography}{10}
\providecommand{\url}[1]{\texttt{#1}}
\providecommand{\urlprefix}{URL }
\expandafter\ifx\csname urlstyle\endcsname\relax
  \providecommand{\doi}[1]{doi:\discretionary{}{}{}#1}\else
  \providecommand{\doi}{doi:\discretionary{}{}{}\begingroup
  \urlstyle{rm}\Url}\fi

\bibitem{wei06:_fast_tcp}
Wei DX, Jin C, Low SH, Hegde S. {FAST} {TCP}: Motivation, architecture,
  algorithms, performance. \emph{{IEEE/ACM} Transactions on Networking}  Dec
  2006; \textbf{14}(6):1246--1259.

\bibitem{jain89}
Jain R. A delay-based approach for congestion avoidance in interconnected
  heterogeneous computer networks. \emph{SIGCOMM Comput. Commun. Rev.}  Oct
  1989; \textbf{19}(5):56--71, \doi{10.1145/74681.74686}.

\bibitem{brakmo94tcp}
Brakmo LS, O'Malley SW, Peterson LL. {TCP} {V}egas: {N}ew techniques for
  congestion detection and avoidance. \emph{SIGCOMM Comput. Commun. Rev.}
  1994; \textbf{24}(4):24--35, \doi{10.1145/190809.190317}.

\bibitem{jabocson90:reno}
Jacobson V. Modified {TCP} congestion avoidance algorithm. email to
  end2end-interest@ISI.EDU mailing list Apr 1990.
  \urlprefix\url{ftp://ftp.ee.lbl.gov/email/vanj.90apr30.txt}.

\bibitem{floyd99:_new_reno_modif_fast_recov_algor}
Floyd S, Henderson T. The new {R}eno modification to {TCP}'s fast recovery
  algorithm. RFC 2582 Apr 1999.
  \urlprefix\url{http://www.ietf.org/rfc/rfc2582.txt}.

\bibitem{martin03}
Martin J, Nilsson A, Rhee I. Delay-based congestion avoidance for {TCP}.
  \emph{{IEEE/ACM} Transactions on Networking}  Jun 2003;
  \textbf{11}(3):356--369.

\bibitem{fu03:_remed_for_perfor_degrad_of}
Fu C, Liew S. A remedy for performance degradation of {TCP} {V}egas in
  asymmetric networks. \emph{{IEEE} Communications Letters}  Jan 2003;
  \textbf{7}(1):42--44.

\bibitem{chan03:_enhan_conges_avoid_mechan_for_tcp_vegas}
Chan YC, Chan CT, Chen YC. An enhanced congestion avoidance mechanism for {TCP}
  {V}egas. \emph{{IEEE} Communications Letters}  Jul 2003;
  \textbf{7}(7):343--345.

\bibitem{liu05:_enhan_tcp_vegas_for_asymm_networ}
Liu J, Chen F, Wei G. Enhanced {TCP} {V}egas for asymmetric networks.
  \emph{Wireless Communications, Networking and Mobile Computing}, vol.~2,
  2005; 1095--1098.

\bibitem{herreria07}
Herrería-Alonso S, Rodríguez-Pérez M, Suárez-González A, Fernández-Veiga
  M, López-García C. Improving {TCP} {V}egas fairness in presence of backward
  traffic. \emph{{IEEE} Communications Letters}  Mar 2007;
  \textbf{11}(3):273--275.

\bibitem{bonal99:_compar_of_tcp_reno_and_tcp_vegasy}
Bonal T. Comparison of {TCP} {R}eno and {TCP} {V}egas: Efficiency and fairness.
  \emph{Performance Evaluation}  Aug 1999; \textbf{36--37}:307--332,
  \doi{10.1016/S0166-5316(99)00037-1}.

\bibitem{weigle06:_perfor_compl_high_speed_tcp_flows}
Weigle MC, Sharma P, Freeman~{IV} JR. Performance of completing high-speed
  {TCP} flows. \emph{Proceedings of {IFIP} {N}etworking 2006}, Coimbra,
  Portugal, 2006; 476--487.

\bibitem{Low02}
Low SH, Peterson L, Wang L. Understanding {V}egas: a duality model. \emph{J.
  ACM}  Mar 2002; \textbf{49}(2):207--235.

\bibitem{Kunniyur03}
Kunniyur S, Srikant R. End-to-end congestion control: utility functions, random
  losses and {ECN} marks. \emph{{IEEE/ACM} Transactions on Networking}  Oct
  2003; \textbf{11}(5):689--702.

\bibitem{tang05:_equil_and_fairn_of_networ}
Tang A, Wang J, Hegde S, Low SH. Equilibrium and fairness of networks shared by
  {TCP} {R}eno and {V}egas/{FAST}. \emph{Telecommunication Systems}  Dec 2005;
  \textbf{30}(4):417--439.

\bibitem{rodriguez08:_heuris_approac_to_passiv_detec}
Rodríguez~Pérez M, Fernández~Veiga M, Herrería~Alonso S, López~García C.
  A heuristic approach to the passive detection of {R}eno-like {TCP} flows.
  \emph{NET-COOP 2008. 2\ensuremath{^{\text{nd}}} Workshop on Network Control
  and Optimization}, \emph{LNCS}, vol. 5425, Altman E, Chaintreau A (eds.),
  Springer-Verlag, 2009; 87--94, \doi{10.1007/978-3-642-00393-6\_11}.

\bibitem{king05:_tcp_afric}
King R, Baraniuk R, Riedi R. {TCP}-{A}frica: {A}n adaptive and fair rapid
  increase rule for scalable {TCP}. \emph{Proceedings of the {IEEE} {INFOCOM}},
  vol.~3, Miami, FL, USA, 2005; 1838--1848.

\bibitem{tang06:_heter_conges_contr}
Tang A, Wei~X D, Low SH. Heterogeneous congestion control: {E}fficiency,
  fairness and design. \emph{{IEEE} International Conference on Network
  Protocols}, Santa Barbara, CA, USA, 2006; 127--136.

\bibitem{hengartner00:_tcp_vegas_revis}
Hengartner U, Bolliger J, Gross T. {TCP} {V}egas revisited. \emph{Proceedings
  of the {IEEE} {INFOCOM}}, vol.~3, Zurich, Switzerland, 2000; 1546--1555.

\bibitem{trinh08:_revis_fast_tcp_fairn}
Trinh TA, Sonkoly B, Molnár S. Revisiting {FAST} {TCP} fairness.
  \emph{18\ensuremath{^{\text{th}}} {ITC} {S}pecialist Seminar on Quality of
  Experience}, Karlskrona, Sweden, 2008.

\bibitem{mo99:_analy_compar_tcp_reno_vegas}
Mo J, La RJ, Anantharam V, Walrand J. Analysis and comparison of {TCP} {R}eno
  and {V}egas. \emph{Proceedings of the {IEEE} {INFOCOM}}, vol.~3, New York,
  NY, USA, 1999; 1556--1563.

\bibitem{la99}
La RJ, Walrand J, Anantharam V. Issues in {TCP} {V}egas Jan 1999.
  Http://www.eecs.berkeley.edu/~ananth/1999-2001/Richard/.

\bibitem{athur01}
Athuraliya S, Li VH, Low SH, Yin Q. {REM}: Active queue management. \emph{IEEE
  Network}  May 2001; \textbf{15}(3):48--53.

\bibitem{chan04}
Chan YC, Chan CT, Chen YC, Ho CY. Performance improvement of congestion
  avoidance mechanism for {TCP} {V}egas. \emph{Proc. {ICPADS}'04}, 2004;
  605--612.

\bibitem{tan05}
Tan L, Yuan C, Zukerman M. {FAST TCP}: fairness and queuing issues.
  \emph{{IEEE} Communications Letters}  Aug 2005; \textbf{9}(8):762--764.

\bibitem{cui06}
Cui T, Andrew L, Zukerman M, Tan L. Improving the fairness of {FAST TCP} to new
  flows. \emph{{IEEE} Communications Letters}  May 2006;
  \textbf{10}(5):414--416.

\bibitem{rodriguez08:_achiev_fair_networ_equil_with}
Rodríguez~Pérez M, Herrería~Alonso S, Fernández~Veiga M, López~García C.
  Achieving fair network equilibria with delay-based congestion control
  algorithms. \emph{{IEEE} Communications Letters}  Jul 2008;
  \textbf{12}(7):535--537.

\bibitem{jacobson88congestion}
Jacobson V. Congestion avoidance and control. \emph{SIGCOMM Comput. Commun.
  Rev.}  1988; \textbf{18}(4):314--329, \doi{10.1145/52325.52356}.

\bibitem{shalunov09:_low_extra_delay_backg_trans_ledbat}
Shalunov S. Low extra delay background transport ({LEDBAT}). IETF Draft Oct
  2009.
  \urlprefix\url{http://tools.ietf.org/html/draft-ietf-ledbat-congestion-00}.

\bibitem{tan06:_compoun_tcp_approac_for_high}
Tan K, Song J, Zhang Q, Sridharan M. A compound {TCP} approach for high-speed
  and long distance networks. \emph{Proceedings of the {IEEE} {INFOCOM}},
  Barcelona, Spain, 2006; 1--12.

\bibitem{ns-2}
NS. {n}s {N}etwork {S}imulator Oct 2005. Http://www.isi.edu/nsman/ns/.

\bibitem{samios03:_model_throug_of_tcp_vegas}
Samios CB, Vernon MK. Modeling the throughput of {TCP} {V}egas.
  \emph{SIGMETRICS Perform. Eval. Rev.}  Jun 2003; \textbf{31}(1):71--81,
  \doi{10.1145/885651.781037}.

\bibitem{alemu04}
Alemu T, Jean-Marie A. Dynamic configuration of {RED} parameters.
  \emph{Proceedings of the {IEEE} {GLOBECOM}}, 2004; 1600--1604.

\bibitem{jin03:_fast_tcp}
Jin C, Wei DX, H~Low S. {FAST TCP}: Motivation, architecture, algorithms,
  performance. \emph{Technical {R}eport}, Caltech CS Dec 17, 2003.

\bibitem{wang05:_model_and_stabil_of_fast_tcp}
Wang J, Wei DX, H~Low S. Modelling and stability of {FAST} {TCP}.
  \emph{Proceedings of the {IEEE} {INFOCOM}}, vol.~2, Pasadena, CA, USA, 2005;
  938--948.

\bibitem{choi05:_global_stabil_of_fast_tcp}
Choi JY, Koo K, Lee JS, Low SH. Global stability of {FAST} {TCP} in single-link
  single-source network. \emph{44th {IEEE} Conference on Decision and Control},
  Seville, Spain, 2005; 1837--1841.

\bibitem{choi06:_global_expon_stabil_of_fast_tcp}
Choi JY, Koo K, Wei DX, Lee JS, Low SH. Global exponential stability of {FAST}
  {TCP}. \emph{45th {IEEE} Conference on Decision and Control}, San Diego, CA,
  USA, 2006; 639--643.

\bibitem{tan07:_param_tunin_for_fast_tcp}
Tan L, Zhang W, Yuan C. On parameter tuning for {FAST} {TCP}. \emph{{IEEE}
  Communications Letters}  May 2007; \textbf{11}(5):458--460.

\bibitem{hasegawa99:_fairn_and_stabil_of_conges}
Hasegawa G, Murata M, Miyahara H. Fairness and stability of congestion control
  mechanisms of {TCP}. \emph{Telecommunications Systems Journal}  Nov 2000;
  \textbf{15}(1--2):167--184.

\bibitem{cardwell04:_tcp_vegas_implem_for_linux}
Cardwell N, Bak B. A {TCP} {V}egas implementation for {L}inux.
  http://neal.nu/uw/linux-vegas/ 2004.

\bibitem{herreria09:_fast_tcp_conges_avoid_implem}
Herrería~Alonso S, Rodríguez~Pérez M. {FAST}-{TCP} congestion avoidance
  implementation Jul 2009.
  \urlprefix\url{http://labredes.det.uvigo.es/~miguel/fast}.

\end{thebibliography}

\end{document}